\documentclass[aps,prb,10pt,twocolumn,reprint,superscriptaddress,floatfix,amsmath,amssymb]{revtex4-2}
\usepackage[utf8]{inputenc}
\usepackage{amsmath}
\usepackage{amssymb}
\usepackage{mathtools}
\usepackage{xcolor}
\usepackage{graphicx}
\usepackage{tabularx}
\usepackage{hyperref}
\usepackage{siunitx}

\newcommand \tr {\operatorname{Tr}}

\newcommand \calO {\mathcal{O}}

\newcommand{\wmax}{\ensuremath{{\omega_\mathrm{max}}}}
\def\mathi{\mathrm i}

\def\deltaF{\delta_{\alpha, \mathrm{F}}}

\newcommand{\chem}[1]{\ensuremath{\mathrm{#1}}}
\newcommand{\tauk}{\ensuremath{\bar{\tau}^\alpha_k}}
\newcommand{\wk}{\ensuremath{\bar{\omega}^\alpha_k}}
\newcommand{\taubar}{\ensuremath{\bar{\tau}}}
\newcommand{\wbar}{\ensuremath{\bar{\omega}}}
\newcommand{\GW}{{\ensuremath{GW}}}
\def\G1{{G^{(1)}}}
\renewcommand{\arraystretch}{1.5}

\begin{document}
\title{
Sparse sampling approach to efficient \textit{ab initio} calculations at finite temperature
}

\author{Jia Li}
\author{Markus Wallerberger}
\affiliation{Department of Physics, University of Michigan, Ann Arbor, MI 48109, USA}
\author{Naoya Chikano}
\affiliation{Department of Physics, Saitama University, Saitama 338-8570, Japan}
\author{Chia-Nan Yeh}
\affiliation{Department of Physics, University of Michigan, Ann Arbor, MI 48109, USA}
\author{Emanuel Gull}
\affiliation{Department of Physics, University of Michigan, Ann Arbor, MI 48109, USA}
\affiliation{Center for Computational Quantum Physics, The Flatiron Institute, New York, New York, 10010, USA}
\author{Hiroshi Shinaoka}
\affiliation{Department of Physics, Saitama University, Saitama 338-8570, Japan}

\begin{abstract}

Efficient \textit{ab initio} calculations of correlated materials at finite temperature require compact representations of the Green's functions both in imaginary time and Matsubara frequency.
In this paper, we introduce a general procedure which generates sparse sampling points in time and frequency from compact orthogonal basis representations, such as Chebyshev polynomials and intermediate representation (IR) basis functions.
These sampling points accurately resolve the information contained in the Green's function,
and efficient transforms between different representations are formulated with minimal loss of information.
As a demonstration,
we apply the sparse sampling scheme to 
diagrammatic \GW{} and GF2 calculations of a hydrogen chain, of noble gas atoms and of a silicon crystal.
\end{abstract}
\maketitle

\section{Introduction}
The finite temperature Green's function formalism~\cite{Matsubara55,abrikosov1975methods,negele1998quantum,luttingerward1960} is a standard approach to study equilibrium properties of correlated quantum systems at finite temperature.
Many-body theories, approximations, and methods based on this formalism are widely used in condensed matter physics, quantum chemistry, and material science.
Applications to low-energy effective model Hamiltonians include lattice Monte Carlo~\cite{blankenbecler1981monte}, dynamical mean-field theory~\cite{georges1996dynamical} with its cluster~\cite{Maier05}, multi-orbital extensions~\cite{Kotliar06,Held06}, and diagrammatic extensions~\cite{Rubtsov:2008cs,Toschi07,Rohringer18}, and diagrammatic or continuous-time quantum Monte Carlo methods~\cite{prokof1998polaron,gull2011continuous}.
In the context of \textit{ab initio} calculations of correlated materials, examples include the \GW{} method~\cite{hedin1965gw,Aryasetiawan_1998,Stan2009,Kutepov2009,GW100,GW100pw,Grumet2018,Kutepov16,Kutepov17}, the self-consistent second order approximation (GF2)~\cite{dahlen2005gf2,phillips2014communication,phillips2015fractional,kananenka2016grid,kananenka2016spline,rusakov2016periodic,welden2016statmech,iskakov2018gf2}, variants of the dynamical mean field theory~\cite{Kotliar06,Sun2002,Biermann02,Tomczak_2012,Tomczak2017,Werner2016}, and the self-energy embedding theory~\cite{Kananenka15,Lan2015communication,Zgid_2017,Lan2017generalized,Lan:2017hk,Tran2018spinunrestricted,Rusakov2019seet}.

The fundamental object of these theories are numerically computed one- and two-particle Green's functions and derived quantities such as self-energies and vertex functions. These quantities are known on a finite Matsubara frequency or imaginary time grid.

In \emph{ab initio} calculations, the difference between the scales of the bare Hamiltonian (which sets a frequency range), temperature (which dictates the frequency resolution), and the energy scales for competing quantum phenomena spans many orders of magnitude.
As a consequence, a naive representation of the Green's function requires an imaginary-time grid too large to store in memory, and solving equations  such as the Dyson equation to the required accuracy becomes prohibitively expensive. This issue becomes even more pronounced for two-particle response functions, which are generically a function of multiple time and orbital indices.

Compact representations of Green's functions are crucial to address this problem.
Representations based on power meshes~\cite{ku2000thesis,Ku2002}, Legendre polynomials~\cite{boehnke2011legendre,kananenka2016grid},  Chebyshev polynomials~\cite{gull2018chebyshev,kutepov2012electronic}, intermediate numerical representations (IR)~\cite{shinaoka2017compressing,chikano2018performance,chikano2019irbasis}, quadrature rules~\cite{steinbeck2000enhancements,kaltak2014low}, and spline interpolations~\cite{kananenka2016spline} have been proposed, as well as high frequency tail expansions~\cite{Rusakov2014,ArminPhD,BluemerPhD}.

Green's function representations, in addition to being compact, also need to enable efficient calculations. The two main stages in most methods are the evaluation of self-energy diagrams (usually best done in imaginary time, as the interaction is instantaneous) and the solution of the Dyson equation (usually best done in Matsubara space, where the equation is diagonal in frequency).
Some of the representations mentioned above are only compact in either time or frequency, and transforming between those domains is expensive or involves a loss of accuracy.
Others, such as the orthogonal polynomial bases, can efficiently and accurately be transformed between coefficients and imaginary time sampling points, but frequency transforms result in a loss of compactness.

It is therefore natural to ask if there is a set of `sparse' sampling points in both frequency and time such that, if the Green's function is evaluated on these points, one may reconstruct the continuous Green's functions
in both time and frequency to high precision. This will then allow to perform diagram calculations in time, Dyson equation solutions in frequency, and transformations in between with minimal loss of accuracy.

This paper presents such a compact representation in both time and frequency by proposing a sparse sampling scheme for finite temperature Green's functions. We illustrate the scheme at the example of Chebyshev~\cite{gull2018chebyshev} and IR basis functions~\cite{shinaoka2017compressing}.
Our method aims to accurately resolve all the information contained within finite temperature Green's functions in a compact set of sampling points, and enables efficient and accurate transforms between imaginary time and Matsubara frequency.

The paper will proceed as follows. In Sec.~\ref{sec:Sparse} we will introduce and derive the sparse sampling method. Sec.~\ref{sec:GF2} will discuss the application to low-order diagrammatic methods such as \GW{} and GF2. Sec.~\ref{sec:Res} will show applications to well-known systems in order to illustrate the power of the method.

\section{Sparse sampling method}\label{sec:Sparse}
\subsection{General description and notation}
The main object of this paper is the Green's function $G^\alpha$, which we assume is expanded into a compact representation in terms of $N$ basis functions, such that in imaginary time and Matsubara frequencies
\begin{align}
    G^\alpha(\tau) &= \sum_{l=0}^{N-1} G^\alpha_l F^\alpha_l(\tau),\label{eq:finite-expansion}\\
    \hat{G}^\alpha(i\omega^\alpha_n) &=  \sum_{l=0}^{N-1} G_l^\alpha \hat{F}^\alpha_{l}(i\omega^\alpha_n)\label{eq:fourier-general} \\
    \hat{F}^\alpha_{l}(i\omega^\alpha_n) &= \int_{0}^\beta d\tau F^\alpha_l(\tau) e^{i\omega^\alpha_n\tau},\label{eq:fourier-coeff}
\end{align}
where $G^\alpha_l$ are expansion coefficients, $F_l^\alpha(\tau)$ imaginary time basis functions with Fourier transform $\hat{F}^\alpha_{l}(i\omega^\alpha_n)$ (the `hat' denoting frequency representations), $\omega^\alpha_n=\pi(2n+\delta_{\alpha,F})/\beta$ Matsubara frequencies, and $\alpha$ denotes the statistics (F for fermions and B for bosons).

In the Chebyshev representation~\cite{gull2018chebyshev}
\begin{align}
    F^\alpha_l(\tau) &\equiv T_l[x(\tau)],
\end{align}
where $T_l(x)$ are Chebyshev polynomials of the first kind and $x(\tau) = 2\tau/\beta-1$.
In the IR basis~\cite{chikano2019irbasis}
\begin{align}
    F^\alpha_l(\tau) &\equiv U_l^\alpha(\tau)
\end{align}
where $U_l^\alpha(\tau)$ depend on the statistics and a dimensionless parameter $\Lambda = \beta\wmax$ with a cutoff frequency $\wmax$.
Appendix~\ref{appendix:compact_basis} summarizes notations for Chebyshev and IR.

To determine $G^\alpha_l$ from $G^\alpha(\tau)$, we choose a finite set of $M$ sampling points $\bar{\tau}^\alpha_0,\ldots,\bar{\tau}^\alpha_{M-1} \in [0, \beta]$ ($M \ge N$).
If $\{\tauk\}$ are chosen such that the discretized basis vectors $\{F^\alpha_0(\tauk)\},\ldots,\{F^\alpha_{N-1}(\tauk)\}$ are linearly independent, the exact values of $G^\alpha_l$ can be computed (transformed) from sampled values of $G^\alpha(\tau)$.
Similarly, if a subset of Matsubara frequencies $\{i\wk\}$ is chosen such that the basis functions are linearly independent,
$G^\alpha_l$ can be obtained from $\hat{G}^\alpha(i\wk)$.

\begin{figure}
    \centering
    \includegraphics[width=.8\linewidth]{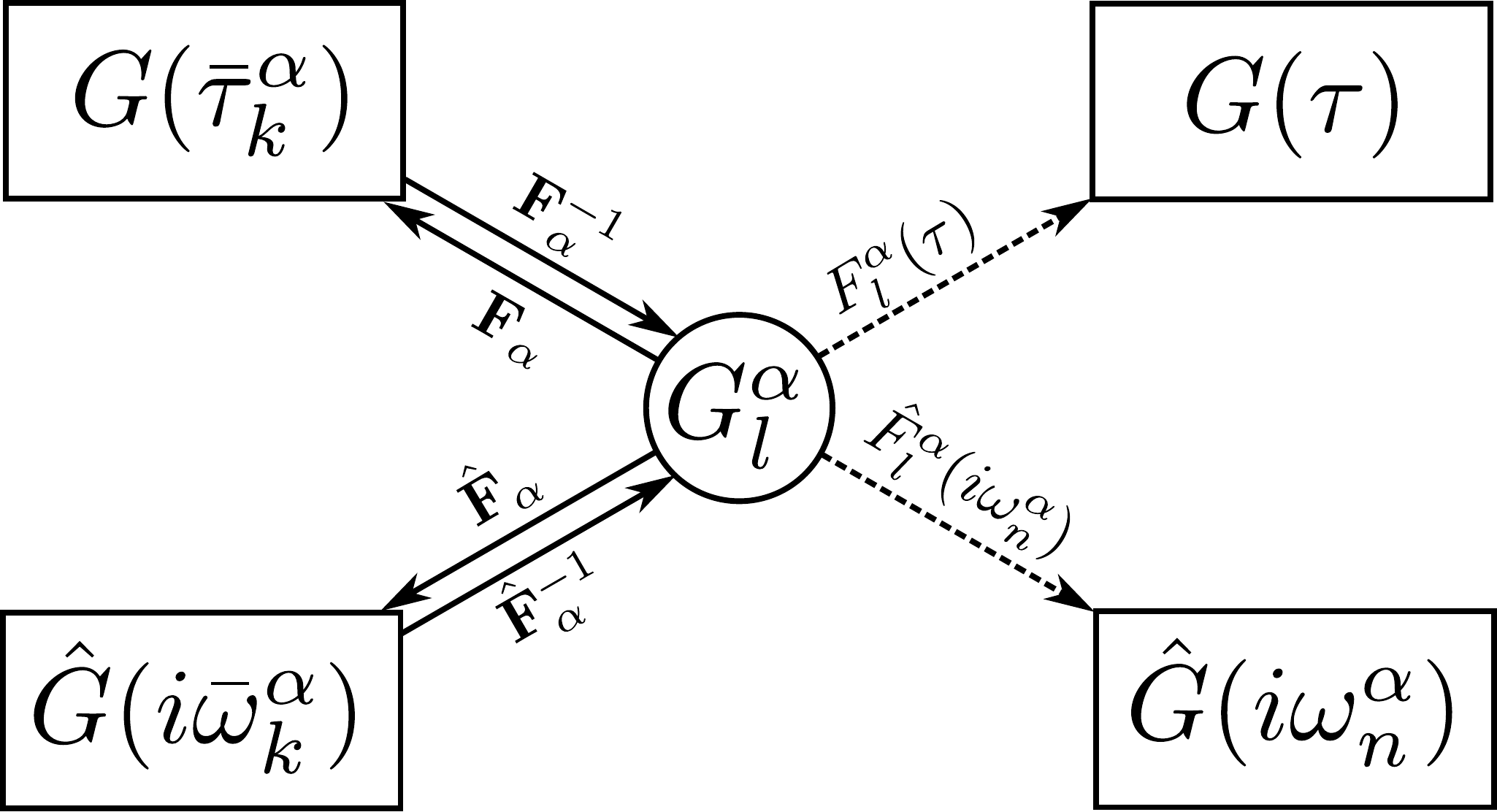}
    \caption{Schematic illustration of relations between different representations.
    Solid lines denote transformations between the basis representation coefficients $G^\alpha_l$ (center) and Green's functions evaluated at imaginary time or frequency sampling points via transformation matrices.
    Dashed lines represent basis expansions of $G^\alpha_l$ to arbitrary imaginary time or frequency points.
    }
    \label{fig:transforms}
\end{figure}

If as many sampling points $M$ are chosen as imaginary time points $N$, these transformations are
\begin{align}
G^\alpha_l &= \sum_{k=0}^{N-1} [\mathbf{F}_\alpha^{-1}]_{lk} G^\alpha(\tauk)\label{eq:back-trans-tau}\\
&= \sum_{k=0}^{N-1} [\hat{\mathbf{F}}_\alpha^{-1}]_{lk} \hat{G}^\alpha(i\wk),\label{eq:back-trans-iw}
\end{align}
where
$\mathbf{F}_\alpha$ and $\hat{\mathbf{F}}_\alpha$ are  $N\times N$ matrices: 
\begin{align}
[\mathbf{F}_\alpha]_{kl} &= F^\alpha_l(\tauk)\label{eq:trans-mat-tau}\\
[\hat{\mathbf{F}}_\alpha]_{kl} &= \hat{F}^\alpha_l(i\wk).\label{eq:trans-mat-iw}
\end{align}
This procedure only requires evaluating the Green's function on $N$ sampling points, and
linear transforms between the time or frequency domain and the basis representation $G^\alpha_l$ become invertible.
$G^\alpha_l$ can thus serve as a proxy to transform between imaginary time and frequency sampling points, as well as evaluation at arbitrary $\tau$ and $i\omega_n$ values, as illustrated in Fig.~\ref{fig:transforms}.

For $M>N$ (more sampling points than basis coefficients), the inverses in Eqs.~(\ref{eq:back-trans-tau}) and (\ref{eq:back-trans-iw}) are replaced by the corresponding pseudoinverses $\mathbf{F}^+ \equiv (\mathbf{F}^\dagger \mathbf{F})^{-1} \mathbf{F}^\dagger$, and the exact transform is replaced by a least squares fitting procedure.

In practical calculations, different choices of basis functions and sampling points lead to differently conditioned equation systems.
A naive choice of sampling points, such as uniformly distributed time or frequency grids, results in almost linearly dependent basis vectors and ill-conditioned transforms, which improve very slowly when additional grid points are added.
For an efficient method, a minimal set of sampling points that generates well-conditioned transformation matrices is desired in order to minimize the number of function evaluations and the  loss of accuracy during transforms.

In the remainder of this section, we show that such a set with $M=N$ can be generated according to the distribution of the roots of the basis functions.

\subsection{Imaginary time sampling}
\textit{Chebyshev}.--
For a truncated Chebyshev basis of size $N$,
the $N$ sampling points in $\tau$ are naturally given by the roots of the $(N+1)$-th basis function $T_{N}(\tau)$ as
\begin{equation}
\tauk \equiv \tau\left(\cos\left(\pi\frac{2k+1}{2N}\right)\right),
\end{equation}
for $k=0,\ldots,N-1$ with the mapping $\tau(x) = \beta (x + 1) / 2$.
These sampling points lead to very well conditioned transformation matrices due to the discrete orthogonality of Chebyshev polynomials, and the condition number of $\mathbf{F}_\alpha$ (defined as $\|\mathbf{F}_\alpha\|_2 \|\mathbf{F}^{-1}_\alpha\|_2$) takes the constant value of $\sqrt{2}$.

\textit{IR basis.}--
For the IR basis with $N$ basis functions and given $\beta$ and $\wmax$, we choose the sampling points $\{\tauk\}$ to be the midpoints of the grid composed of the $N-1$ roots of the highest order basis function $U_{N-1}^\alpha(\tau)$ and the boundary points 0 and $\beta$.
We choose not to use the roots of the next basis function $U^\alpha_N(\tau)$ like in the Chebyshev case due to the fact that the IR basis is a numerical basis, and thus it is more convenient to determine sampling points from the available basis functions.

\subsection{Matsubara frequency sampling}
We generate sampling points in Matsubara frequencies following an algorithm similar to the imaginary time cases, with two additional considerations.
First, function values should only be evaluated on the discrete Matsubara frequencies.
Second, fermionic and bosonic Matsubara frequencies have to be treated separately, and the zero bosonic frequency (which represents static physics) has to be considered explicitly.

\textit{Chebyshev.}--In the Chebyshev representation, we follow the same heuristics as in the $\tau$ sampling by finding or approximating zeros of the next basis function $\hat{T}^\alpha_{N}(i\omega^\alpha_n)$ defined in Matsubara frequency.

For fermions and even $N$, when continued to continuous Matsubara frequency space, $\hat{T}^\mathrm{F}_{N}(i\omega^\alpha_n)$ has $N$ roots on the imaginary axis $(-i\infty, i\infty)$.
We take $N$ Matsubara frequencies closest to these roots as sampling points.

For bosons and odd $N$, $\hat{T}^\mathrm{B}_{N}(i\omega^\alpha_n)$ has $N-1$ roots.
We define $N-1$ sampling points as the  Matsubara frequencies closest to the roots.
We take the zero bosonic frequency $i\omega^\mathrm{B}_n = 0$ as the last sampling point.
The zero bosonic frequency, which corresponds to a constant offset in $\tau$
and often has to be treated separately, serves as a natural complement.

The requirement that even $N$ should be used for fermions and odd $N$ for bosons is necessary because the other cases (odd $N$ for fermions or even $N$ for bosons) will not yield adequate number of sampling points due to the analytic structure of $\hat{T}^\alpha_N(i\omega^\alpha_n)$. See Appendix~\ref{appendix:sample-detail} for a discussion.

\textit{IR basis.}--For the IR basis, the procedure for getting frequency sampling points is more empirical due to the numerical nature of the basis function.
We partition all Matsubara frequencies into contiguous groups such that the highest order basis function $\hat{U}^\alpha_{l_\mathrm{max}}(i\omega^\alpha_n)$ has the same sign within each group ($l_\mathrm{max} \equiv N-1$).
$\hat{U}^\alpha_{l_\mathrm{max}}(i\omega^\alpha_n)$ is either purely real (for even $l_\mathrm{max}$) or purely imaginary (for odd $l_\mathrm{max}$).
We therefore use the sign of the corresponding real or imaginary part as the sign of $\hat{U}^\alpha_{l_\mathrm{max}}(i\omega^\alpha_n)$.
The sampling points $i\wk$ are chosen to be those that maximize $|\hat{U}^\alpha_{l_\mathrm{max}}(i\omega^\alpha_n)|$ in each group.

By checking the resulting sampling points numerically, we conclude that by requiring $N$ to be even ($l_\mathrm{max}$ odd) for fermionic basis and odd ($l_\mathrm{max}$ even) for bosonic basis, the number of sampling points is exactly $N$, and the bosonic sampling points naturally include zero.

\subsection{Numerical demonstration}

\begin{figure}
    \centering
    \includegraphics[width=\linewidth,trim=0 8.5cm 0 0, clip]{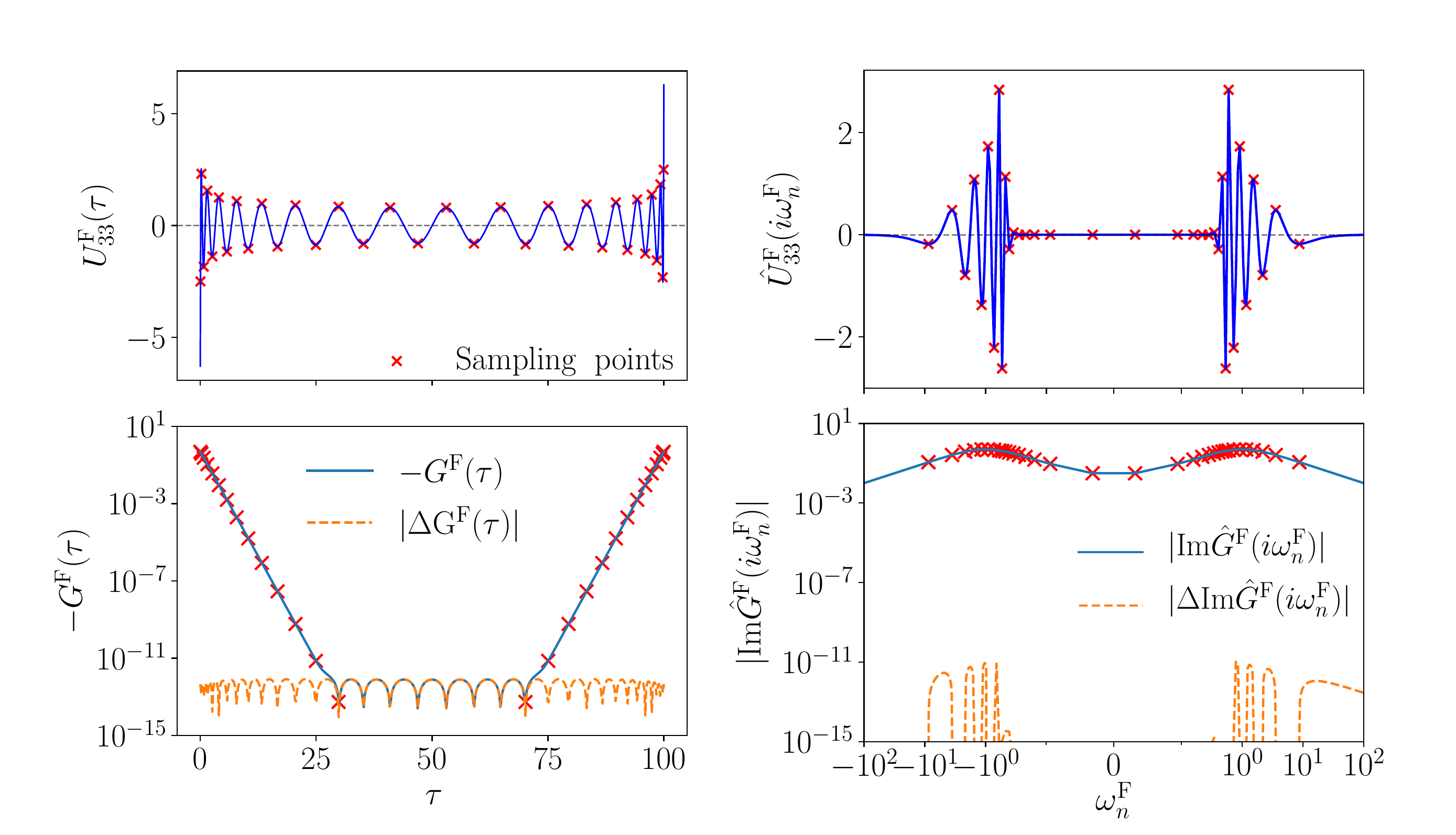}
    \caption{
    Distribution of sampling points and results transformed from imaginary time (left panels) and Matsubara frequency (right panels) for the IR basis by sparse sampling.
    We consider a model of a semicircular density of states of half width 1 at $\beta=100$ defined in Eq.~(\ref{eq:semicirc-dos}).
    We take $\wmax=1$ for the IR basis.
    The sampling points are denoted by crosses.
    Top row: The basis functions used to generate sampling points ($l=33$).
    Bottom row: Comparison of the reconstructed Green's function to exact results.
    }
    \label{fig:sampling}
\end{figure}

The transformation defined in Eqs.~(\ref{eq:back-trans-tau}) and (\ref{eq:back-trans-iw}) is exact if the Green's function is a linear combination of a finite set of basis functions (\ref{eq:finite-expansion}).  With physical Green's functions, this is seldom the case, and any finite expansion incurs a truncation error.  Fortunately, in both the IR and the Chebyshev expansion, the truncation error is controlled: the analyticity of the finite-temperature Green's function in $(0,\beta)$ guarantees exponential convergence of the Chebyshev expansion, and the construction of the IR basis from analytic continuation guarantees the same thing for the IR expansion~\cite{shinaoka2017compressing}.

To demonstrate the behavior of the sparse sampling scheme when applied to physical Green's functions,
we consider a model with semicircular density of states (full bandwidth is $2$) for the IR basis in Fig.~\ref{fig:sampling}:
\begin{align}
    \rho(\omega) &= \frac{2}{\pi}\sqrt{1-\omega^2}.\label{eq:semicirc-dos}
\end{align}

As an example, the top left panel shows $U_{34}^{\mathrm{F}}$ as a function of imaginary time (blue lines) and illustrates the $N=34$ sampling points for the fermionic basis of $\beta=100$ and $\wmax=1$ (red crosses).
The sampling points cluster near $\tau=0$ and $\beta$, where this basis function is rapidly oscillating.

The top right panel of Fig.~\ref{fig:sampling} shows the distribution of the Matsubara frequency sampling points
generated for the same basis for $N=34$.
The Fourier transformed basis function $\hat{U}_{33}^{\mathrm{F}}(i\omega_n)$ (blue lines) exhibits $N-1=33$ sign changes, which define $N=34$ sampling points (red crosses).
The sampling points are distributed almost logarithmically, which allows us to capture all the features of $\hat{U}^\mathrm{F}_{l_\mathrm{max}}(i\omega_n)$ from low to high frequency.

In the bottom row of Fig.~\ref{fig:sampling}, the left and right panels
illustrate the sampling of $G(\tau)$ and $\hat{G}(i\omega_n)$, respectively.
The sampling points capture relevant features of the Green's function in both cases.
We compare the interpolated and extrapolated results with the numerically exact values.
For imaginary time, one can see agreement at the level of $\sim 10^{-12}$ in the whole interval of $[0, \beta]$,
which matches the singular value cutoff we used ($10^{-12}$).
For Matsubara frequency, the coefficients obtained by the sparse sampling not only interpolate $\hat{G}(i\omega_n)$ but also extrapolate it precisely beyond the highest sampling frequency.

\subsection{Technical details}
In practical applications, it is advisable to precompute the sampling points and transformation matrices for the basis functions employed, to avoid unnecessary evaluations of $\hat{T}^\alpha_l(i\omega^\alpha_n)$ and $\hat{U}^\alpha_l(i\omega^\alpha_n)$.

The \texttt{irbasis} library~\cite{chikano2019irbasis} provides numerical data of the IR basis functions in the dimensionless form
for selected values of $\Lambda$ from $\Lambda=10$ up to $\Lambda = 10^7$.
The numerical evaluation of the basis functions in Matsubara frequency is also implemented.

The procedures we have presented are not unique, and we do not claim that they are optimal definitions of sampling points.
One may design other choices with similar or, potentially, better numerical performance.
The number of sampling points may also exceed the basis size $N$, as long as inversions in Eqs.~(\ref{eq:back-trans-tau}) and (\ref{eq:back-trans-iw}) are replaced with a pseudoinverse, and the resulting transformations are well-conditioned.
Nevertheless, the algorithms introduced in this section provide a systematic and unambiguous way to obtain the minimum sets of sampling points which yield well-conditioned numerical transforms and high accuracy.
Appendix~\ref{appendix:sample-detail} includes numerical investigations of the transformation matrices.

\section{Sparse sampling approach to solving diagrammatic equations}
\label{sec:GF2}

\begin{figure*}
    \centering
    \includegraphics[width=\textwidth]{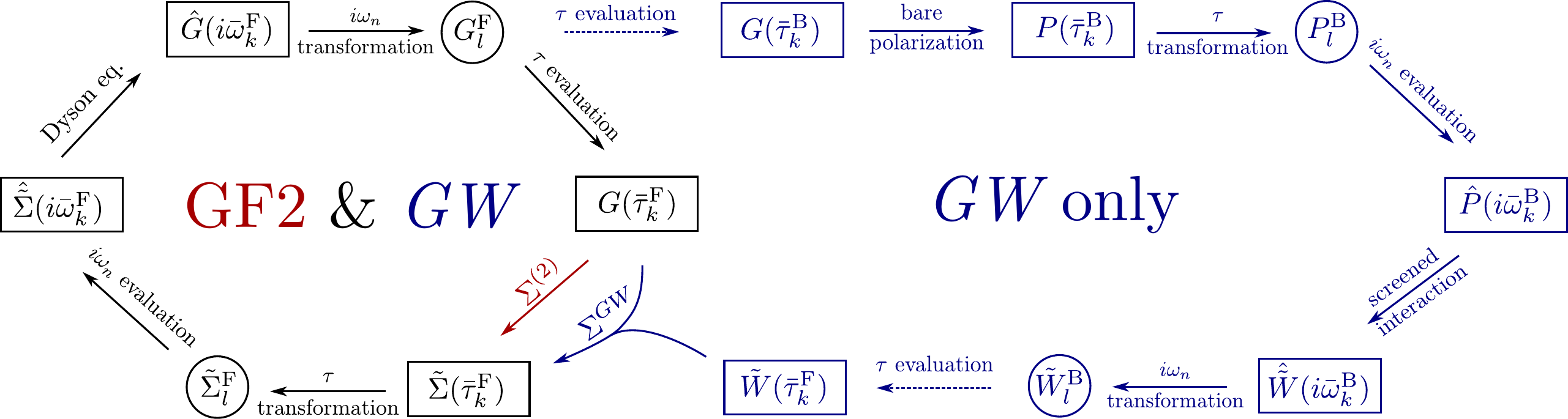}
    \caption{
    Illustration of GF2 and \GW{} procedures using the sparse sampling scheme.
    The red and blue lines denote GF2-only and \GW-only steps, respectively.
    Dashed arrows indicate evaluations that change the statistics of the representation.
    }
    \label{fig:gf2-procedure}
\end{figure*}

Self-consistent second order Green's function theory (GF2) is a second order perturbation theory which renormalizes the Green's function~\cite{dahlen2005gf2,phillips2014communication,phillips2015fractional,kananenka2016grid,kananenka2016spline,rusakov2016periodic,welden2016statmech,iskakov2018gf2}.
Self-consistent \GW{}~\cite{hedin1965gw,Stan2009,Grumet2018,Kutepov16,Kutepov17} further renormalizes the interaction.

GF2 and \GW{} calculations involve evaluations of diagrams in $\tau$ and solving convolution equations in Matsubara frequencies.
Quantities such as the Green's function and the screened interaction are repeatedly converted back and forth between time and frequency.
These calculations therefore serve as a good test platform to demonstrate the numerical stability and computational advantage of the sparse sampling scheme.

In \textit{ab initio} calculations of molecules, a finite spatial basis set
is chosen for each atom.
Since the energy scale $\wmax$ in IR basis is only well defined for Green's functions under orthogonal spatial orbitals, for better estimations of $\wmax$ or $\Lambda$
we perform canonical orthogonalizations~\cite{ostlund1996modern} on the atomic orbitals for a chosen basis set in each system, and use the resulting orthogonal spatial basis in all calculations.
The Hamiltonian reads
\begin{equation}
    H = \sum_{ij\sigma} h_{ij} c^\dagger_{i\sigma} c_{j\sigma} +
    \frac{1}{2}\sum_{ijkl}\sum_{\sigma\sigma'} V_{ijkl} c^\dagger_{i\sigma} c^\dagger_{k\sigma'} c_{l\sigma'} c_{j\sigma}\label{eq:hamiltonian}
\end{equation}
where $h$ is the single-particle Hamiltonian, and $V$ is the Coulomb integrals defined on the orthogonalized orbitals.

The self-consistency in GF2 and \GW{} is defined by the Dyson equation
\begin{align}
    \hat{G}(i\omega^{\mathrm{F}}_n) &= [(i\omega^{\mathrm{F}}_n + \mu)I - F - \hat{\tilde{\Sigma}}(i\omega^{\mathrm{F}}_n)]^{-1}.\label{eq:dyson}
\end{align}
The Fock matrix $F = h + \Sigma^{\mathrm{HF}}$ includes the frequency independent Hartree-Fock contribution
\begin{equation}
    \Sigma^{\mathrm{HF}}_{ij}=(2V_{ijkl} - V_{ilkj})\rho_{kl},
\end{equation}
where $i, j, k, l$ are orbital indices.
GF2 approximates the frequency-dependent self-energy $\tilde{\Sigma}$ as
\begin{align}
    \tilde{\Sigma}_{ij}(\tau) &= -G_{kl}(\tau)G_{qm}(\tau)G_{np}(-\tau)\times \nonumber\\
    &\times V_{ikpq} (2V_{ljmn} - V_{mjln}),\label{eq:sigma2}
\end{align}
while \GW{} approximates $\tilde{\Sigma}$ as
\begin{equation}
    \tilde{\Sigma}_{ij}(\tau) = -G_{lk}(\tau) \tilde{W}_{ilkj}(\tau),\label{eq:GW}
\end{equation}
where $\tilde{W} = W - V$ is the frequency-dependent part of the screened interaction $W$.
$W$ is calculated by the random phase approximation (RPA)~\cite{PhysRev.92.609} as
\begin{equation}
    \hat{W}_{ijkl}(i\omega^{\mathrm{B}}_n) = V_{ijkl} + V_{ijpq} \hat{P}_{qpsr}(i\omega^{\mathrm{B}}_n) \hat{W}_{rskl}(i\omega^{\mathrm{B}}_n),\label{eq:screened-inter}
\end{equation}
where the bare polarization is given by 
\begin{equation}
    P_{ijkl}(\tau) = -G_{il}(\tau) G_{jk}(-\tau).\label{eq:polarization}
\end{equation}

At self-consistency, physical properties are evaluated from $G$ and $\Sigma$. For example, density matrix $\rho$ is given by $\rho_{ij} = G_{ji}(0^-) = -G_{ji}(\beta)$, and the total electronic energy is
\begin{align}
    E =& \tr{[\rho H_0]} + \frac{1}{2}\tr{[\rho \Sigma_{\mathrm{HF}}]} \nonumber\\
    &+ \frac{1}{2\beta}\sum_{n} \tr{[\hat{\tilde{\Sigma}}(i\omega^{\mathrm{F}}_n) \hat{G}(i\omega^{\mathrm{F}}_n)]}.\label{eq:etotal}
\end{align}

Fig.~\ref{fig:gf2-procedure} illustrates how the self-consistent calculations can be performed by sparse sampling.
In a GF2 calculation, we first evaluate Green's function at sampling points $\taubar^{\mathrm{F}}_k$, and construct the self-energy $\tilde{\Sigma}(\taubar^{\mathrm{F}}_k)$ following second-order approximation (\ref{eq:sigma2}).
The self-energy is then transformed to the basis representation $\tilde{\Sigma}^{\mathrm{F}}_l$ following Eq.~(\ref{eq:back-trans-tau}), which is then evaluated on the frequency sampling points to get $\hat{\tilde{\Sigma}}(i\wbar^{\mathrm{F}}_k)$.
The Dyson equation (\ref{eq:dyson}) is solved for each $i\wbar^{\mathrm{F}}_k$ to obtain $\hat{G}(i\wbar^{\mathrm{F}}_k)$.
We then transform the Green's function to its basis representation $G^{\mathrm{F}}_l$ following Eq.~(\ref{eq:back-trans-iw}).
The updated Green's function in $\tau$ is recovered by evaluating $G^{\mathrm{F}}_l$  on sampling points $\taubar^{\mathrm{F}}_k$.
The procedure is repeated using the updated Green's function until self-consistency, which corresponds to the inner loop in Fig.~\ref{fig:gf2-procedure}.
We can see that the compact basis representations $G^{\mathrm{F}}_l$ and $\tilde{\Sigma}^{\mathrm{F}}_l$ serve as proxies to convert back and forth between $\tau$ and frequency domains, all evaluated on corresponding sampling points.

While all quantities involved in GF2 are fermionic, in \GW{} we have to switch between fermionic quantities ($G$ and $\Sigma$) and bosonic quantities ($P$ and $W$).
This is achieved again by using compact basis representations as proxies: when calculating the polarization $P$, we evaluate $G^{\mathrm{F}}_l$ on the \textit{bosonic} sampling points $\taubar^{\mathrm{B}}_k$ and assemble $P(\taubar^{\mathrm{B}}_k)$ following Eq.~(\ref{eq:polarization}).
We then carry out the calculation of $\tilde{W}$ on the frequency sampling points $i\wbar^{\mathrm{B}}_k$, and obtain the compact basis representation $\tilde{W}^{\mathrm{B}}_l$.
Finally we evaluate $\tilde{W}^{\mathrm{B}}_l$ back on the \textit{fermionic} sampling points $\taubar^{\mathrm{F}}_k$, and compute self-energy using the \GW{} approximation (\ref{eq:GW}).
The rest of the procedure is identical to GF2. 

Collective physical observables such as the total energy and the density matrix can be evaluated accurately from the compact basis representation of $G$ and $\Sigma$.
Efficient evaluation of the density is beneficial in calculations with a fixed number of electrons, where one has to adjust the chemical potential in each self-consistent iteration~\cite{phillips2014communication}.
Appendix~\ref{appendix:self-consistent-detail} discusses these technical aspects in detail.

Note that since most basis functions, including Chebyshev and IR, cannot capture constant shifts in frequency (which correspond to a delta function at $\tau=0$), it is important that one only expands the frequency-dependent components such as $\tilde{\Sigma}$ and $\tilde{W}$ using compact basis representations.
One also needs to be careful when zero-energy mode exists in a bosonic quantity, in which case the IR basis function cannot describe the constant shift in imaginary time~\cite{chikano2018performance}.

\section{Results}\label{sec:Res}
\subsection{Hydrogen chain}

We first apply our sparse sampling scheme to GF2 and \GW{} caculations of a system composed of 10 hydrogen atoms placed on a straight line with equal spacing $r$.
The hydrogen chain, due to its simplicity, serves as a benchmark platform for testing numerical methods of correlated electrons.
Reference data for the hydrogen chain were carefully compared and analyzed in Ref.~\onlinecite{motta2017hydrogen} with many methods including GF2 and \GW.
It is therefore convenient to use this system to analyze the sparse sampling scheme.

We perform GF2 and \GW{} calculations for \chem{H_{10}} with $r = \SI{1}{\bohr}$  and $\beta=\SI{1000}{\hartree^{-1}}$ ($T\sim \SI{315.8}{\kelvin}$).
We use the minimal basis set STO-6g with only one 1$s$ orbital per atom.
Hartree-Fock calculations show that the difference between the highest and the lowest Hartree-Fock energy levels is about $\Delta E \approx \SI{5.76}{\hartree}$ ($\sim \SI{156}{\electronvolt}$).
The dimensionless parameter for the IR basis can thus be estimated by taking $\Lambda$ to bound the value $\beta\Delta E \approx 5.76\times 10^3$.
We take $\Lambda = 10^5$ in all our calculations.

Both Chebyshev and IR basis functions are used together with the sparse sampling scheme.
To demonstrate convergence, we examine a series of calculations with different sizes of basis functions.
Typically we choose $N$ to be an even number, which is then used as the size of the fermionic basis.
For corresponding bosonic basis functions in \GW, we used the closest odd number $N-1$ as the basis size.
The Python library \texttt{irbasis} version 2.0.0b1~\cite{chikano2019irbasis} is used for calculating IR basis functions.

With each basis of fixed size $N$, we perform self-consistent GF2 and \GW{} calculations following the procedures described in the previous section.
The initial guess of the Green's function is constructed from Hartree-Fock calculations using the PySCF library~\cite{PYSCF}.
Self-consistent calculations, as illustrated in Fig.~\ref{fig:gf2-procedure}, are then executed repeatedly until the energy difference is below the convergence tolerance $E_\mathrm{tol} = \SI{e-8}{\hartree}$ between two consecutive iterations.
Energy values converged with respect to basis size $N$ are cross-checked with reference data to ensure correctness.
All results are then compared to converged values to assess errors in energy and density matrix as a function of basis size $N$.

\begin{figure}
    \centering
    \includegraphics[width=\linewidth]{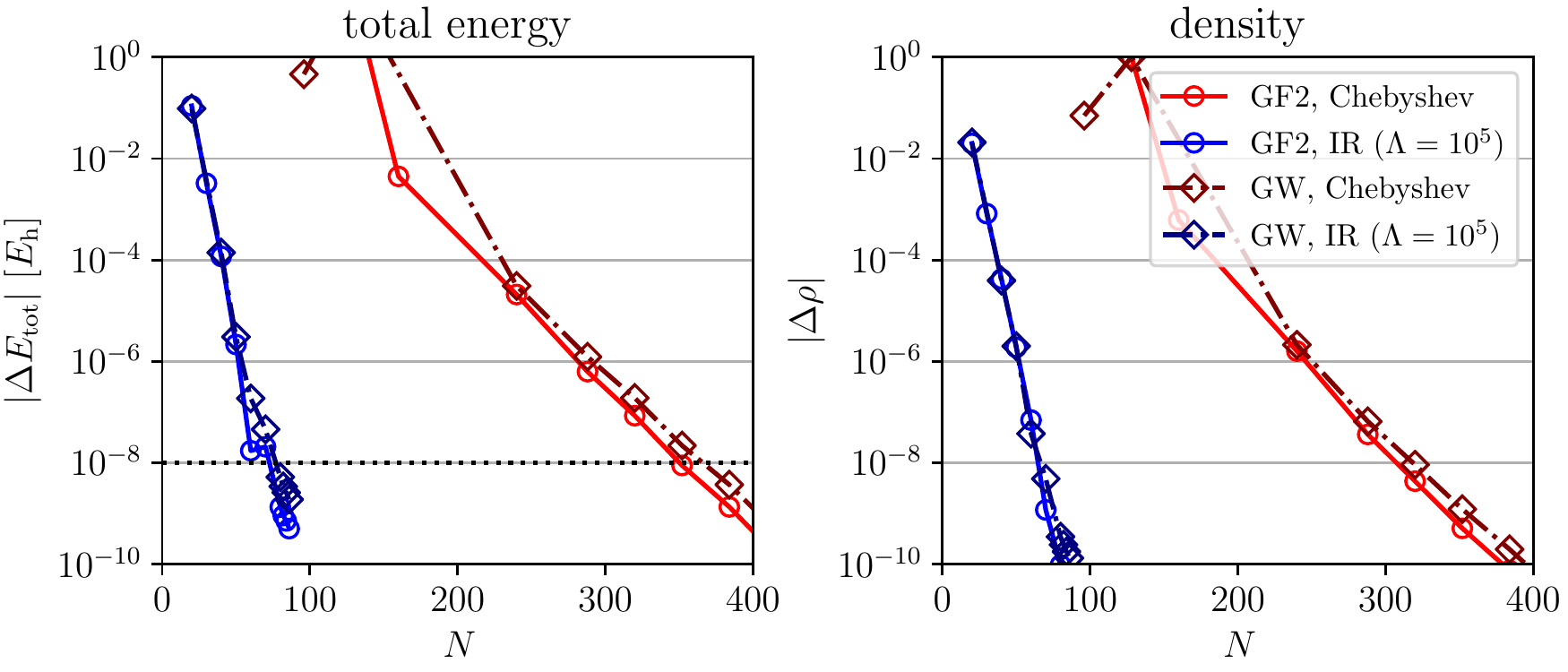}
    \caption{Error in total energy and density from converged GF2 and \GW{} calculations for H${}_{10}$ in minimal basis at $\beta=\SI{e3}{\per\hartree}$. The left panel shows the convergence of the total energy with the dashed horizontal line representing the convergence threshold of \SI{e-8}{\hartree}. The right panel shows the convergence of the density.
    \label{fig:etot}}
\end{figure}

In the left panel of Fig.~\ref{fig:etot}, we show the convergence of errors in total energy with respect to $N$.
GF2 and \GW{} share similar convergence behavior, in that the error decreases almost exponentially with respect to $N$ in either basis representation.
In the Chebyshev representation, we obtain convergence of the total energy below the tolerance $E_\mathrm{tol}$ with around 350 Chebyshev polynomials.
With IR basis, convergence is reached with less than 100 basis functions.
Similarly, the right panel of Fig.~\ref{fig:etot} illustrates the convergence of the maximum error in density matrix, exhibiting an exponential decay of errors.
This indicates that with a reasonable number of sampling points, we can reach very high precision in both global observables such as the energy and local properties like the density matrix.
The observation that \GW{} shows a convergence behavior similar to GF2 indicates that no substantial additional errors are introduced during the frequent switching between fermionic and bosonic representations.

\begin{figure}
    \centering
    \includegraphics[width=\linewidth]{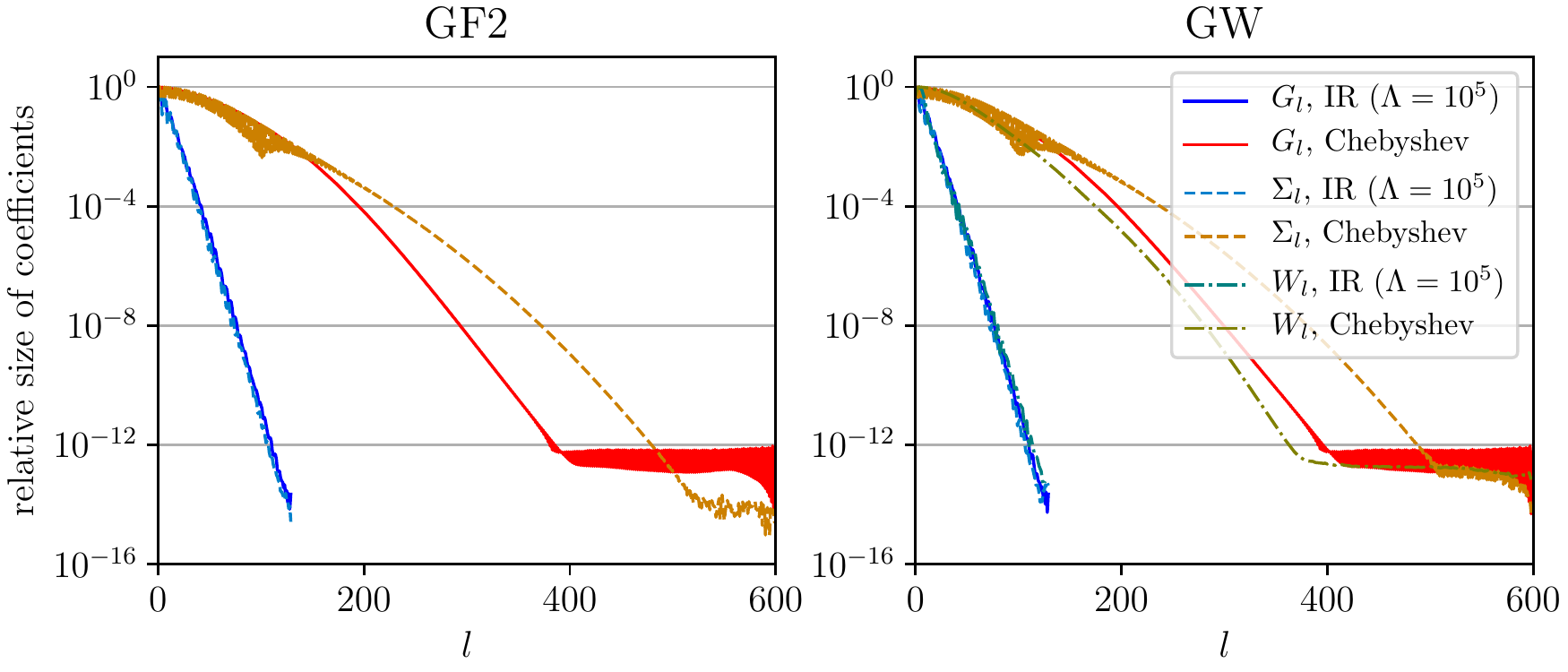}
    \caption{Relative size of basis expansion coefficients with converged GF2 (left panel) and \GW{} (right panel) calculations for $\chem{H_{10}}$ in STO-6g at $r=1.0$ a.u., $\beta=\SI{1000}{\per\hartree}$.
    }
    \label{fig:dyson_compare}
\end{figure}

The sparse sampling scheme is stable
thanks to the well-conditioned transformation matrices generated from the sampling points.
We demonstrate this in Fig.~\ref{fig:dyson_compare}, which shows the relative magnitude of basis expansion coefficients for the converged solution.
$N$ is chosen large enough to ensure that all quantities are well approximated by the basis representations, with $N=600$ for Chebyshev and $N=130$ for IR with $\Lambda = 10^5$.
Even after several iterations with multiple transforms forward and backward between different types of sampling points, we see that for all quantities, expansion coefficients decay at least exponentially as we expect from the properties of the basis, down to a relative size below $10^{-12}$.
The truncation error due to the finite basis expansion is therefore controlled, and no amplification of error is observed during the self-consistent iterations.

The sparse sampling scheme ensures that the number of $\tau$ grid points and the number of Matsubara frequencies is the same as the basis size $N$.
We reach a precision of 8 digits in total energy with only hundreds of $\tau$ points and Matsubara frequencies, a significant improvement from the conventional approach used in Ref.~\onlinecite{motta2017hydrogen}, where $\sim 10^4$ Matsubara frequencies were used for higher temperature ($\beta \sim \SI{100}{\hartree^{-1}}$) and bigger convergence threshold ($E_\mathrm{tol}\sim$ $\SI{e-6}{\hartree}$ or $\SI{e-7}{\hartree}$).
This greatly reduces the computational cost and memory requirement in all parts of the calculations while still being accurate.
The sparse sampling scheme thus allows to tackle problems that were too costly to calculate, especially low temperature calculations of systems with large energy scales.

\subsection{Noble gas atoms}
Noble gas atoms such as Kr have deep core states, if no pseudopotentials are employed.
Due to the large energy scale caused by the core states, it is computationally very demanding for conventional methods to resolve sharp features close to $\tau=0$ or $\beta$ in $G(\tau)$ or the slow decay of $\hat{G}(i\omega_n)$ at high frequency.
Even with a compact polynomial basis such as Chebyshev, thousands of basis functions are required to represent the Green's functions~\cite{gull2018chebyshev}, and
effective core potentials (ECP), which absorb electron in inner orbitals to the ionic potential, have to be employed in most practical calculations.
We choose this problem to demonstrate the power of the sparse sampling method when dealing with large energy scales, while avoiding additional physical or technical difficulties.

The IR basis is a natural choice for systems with large energy scales. As long as the spectral cutoff $\wmax$ is chosen to include all energy scales in the system, exponential convergence of the coefficients is guaranteed by construction, usually with no more than a couple hundred basis functions~\cite{shinaoka2017compressing}.
The full potential of IR basis is realized when combined with the sparse sampling scheme developed in this paper.
Numerical difficulties in either $\tau$ or frequency domain are reduced to a single issue: whether the basis functions can capture all relevant quantities well.
Therefore, by using a sparse sampling method with the IR basis, we are able to treat noble gas atoms efficiently even in all-electron calculations.

We choose the all-electron correlation consistent basis set cc-pVDZ~\cite{Dunning1989}, and perform GF2 and \GW{} calculations of four noble gas atoms: He, Ne, Ar, and Kr, at $\beta=\SI{1000}{\hartree^{-1}}$.
Similar to the case for \chem{H_{10}}, we estimate the dimensionless parameter $\Lambda$ according to the Hartree-Fock energy spectrum for each individual atom: $\Lambda = 10^4$ for He, $\Lambda = 10^5$ for Ne, and $\Lambda = 10^6$ for Ar and Kr.
The sparse sampling scheme is used in all calculations, and we vary the basis size $N$ to explore the convergence behavior.
The energy convergence threshold is set to $E_\mathrm{tol} = \SI{e-8}{\hartree}$, which is much smaller than the energy scale of Kr ($\sim \SI{e3}{\hartree}$).

\begin{figure}
    \centering
    \includegraphics[width=\linewidth]{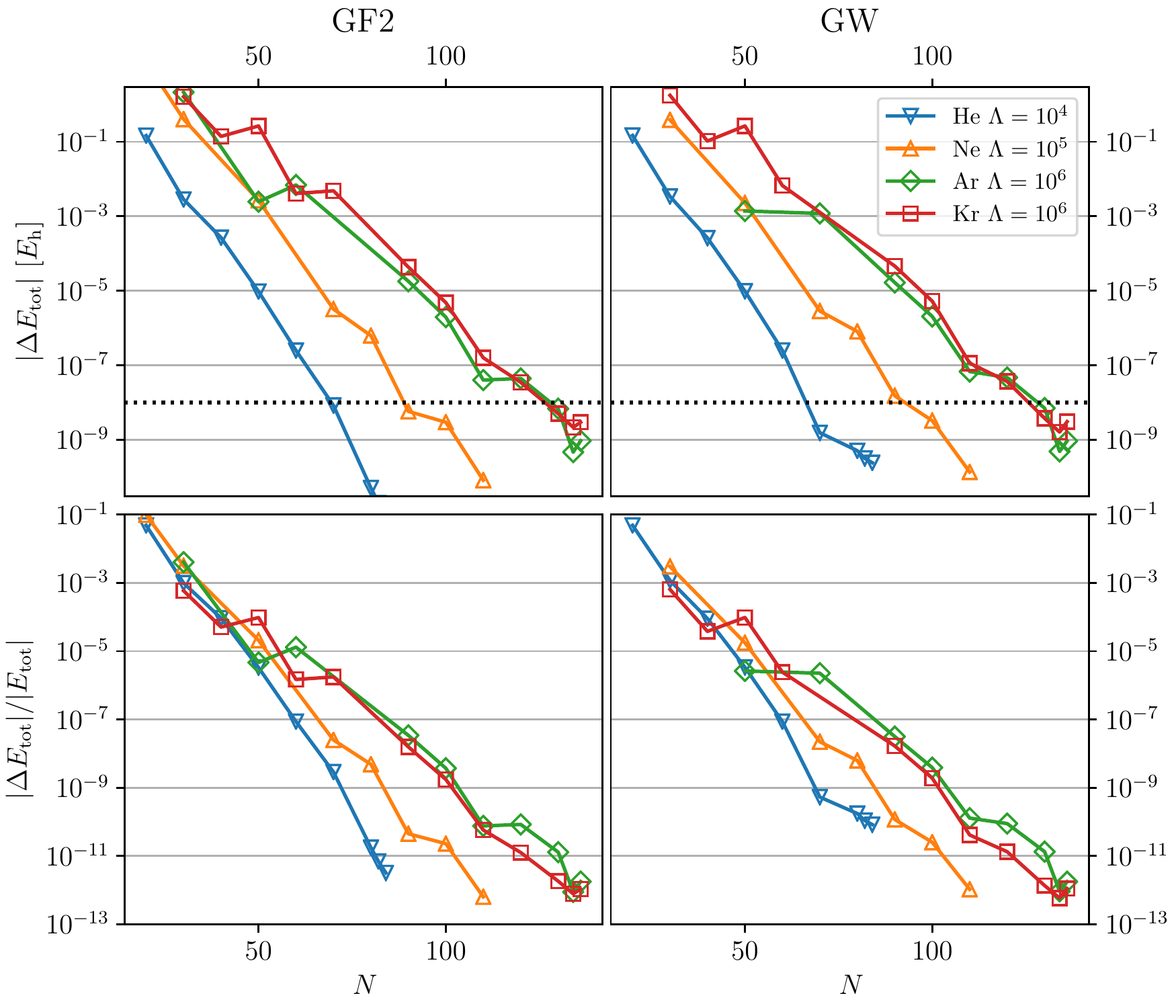}
    \caption{Total energy convergence in GF2 and \GW{} calculations of noble gas atoms with respect to the number of IR basis functions $N$.
    Left and right columns show GF2 and \GW{} results respectively.
    Top row: absolute differences in total energy to the converged value.
    Dashed black horizontal line illustrates the energy convergence threshold of $\SI{e-8}{\hartree}$.
    Bottom row: relative differences in total energy.
    }
    \label{fig:nobel-gas}
\end{figure}

Fig.~\ref{fig:nobel-gas} shows the energy convergence of all four atoms with respect to basis size $N$ with GF2 (left column) and \GW{} (right column).
The upper panels indicate that the basis converges for all atoms, with absolute difference in energy dropping in an exponential trend below the convergence tolerance $E_\mathrm{tol}$.
The lower panels put the convergence in a relative scale, where all atoms in both GF2 and \GW{} can reach $\sim 10^{-10}$ relative convergence.

Our results show that we can obtain fast basis convergence with around 100 basis functions for all systems studied.
This is consistent with the property of IR basis that the basis size $N$ scales only logarithmically with $\Lambda$~\cite{chikano2018performance}.

\subsection{Application to solids}

One can apply the sparse sampling technique to periodic systems the same way as in molecules.
Additional momentum dependence can be viewed as an extra ``orbital index''.
As an example, we apply our method to the \GW{} calculation of a silicon crystal, using a $4\times 4\times 4$ momentum mesh, inverse temperature $\beta=\SI{1000}{\hartree^{-1}}$, and energy self-consistency threshold $E_\mathrm{tol} = \SI{e-6}{\hartree}$.
We choose the GTH-DZVP basis of Gaussian orbitals~\cite{Goedecker1996}, which has 13 orbitals per Si atom.
The initial Green's function is constructed from the LDA solution in this basis set with a GTH-LDA pseudopotential.
We orthogonalize the spatial basis set in all \GW{} calculations.
The LDA spectrum shows a maximum energy scale of $\Delta E \approx \SI{3}{\hartree}$, thus we choose $\Lambda=10^4$ and $10^5$ for IR basis to bound the value $\beta\Delta E \approx 3\times 10^3$ .

Table~\ref{tab:silicon-etot} shows the total energy convergence with respect to the size of basis representations $N$.
With both Chebyshev and IR, the total energy per unit cell converges below the self-consistency threshold of $\SI{e-6}{\hartree}$ to the same value $E_\mathrm{tot}=\SI{-7.880430}{\hartree}$ as $N$ increases.
The Chebyshev basis converges at around $N\approx 300$, and the IR basis at $N\approx 80$ for $\Lambda=10^4$, $N\approx 110$ for $\Lambda=10^5$.
In contrast, conventional approaches using uniform Matsubara frequency grid and high frequency tail expansions require $\sim 10^5$ frequency points.

\begin{table}
    \centering
    \caption{\GW{} total energy of the silicon crystal per unit cell at $\beta=\SI{1000}{\per\hartree}$, with self-consistency convergence threshold $E_\mathrm{tol}=\SI{e-6}{\hartree}$.}
    \label{tab:silicon-etot}
    \begin{ruledtabular}
    \begin{tabular}{l l l l l l}
        \multicolumn{2}{c}{Chebyshev} &
        \multicolumn{2}{c}{IR $\Lambda=10^4$} &
        \multicolumn{2}{c}{IR $\Lambda=10^5$}\\
        \hline
        $N$ & $E_\mathrm{tot}$~[\si{\hartree}] &
        $N$ & $E_\mathrm{tot}$~[\si{\hartree}] &
        $N$ & $E_\mathrm{tot}$~[\si{\hartree}] \\
        \hline
        100 & -8.0270874 & 80 & -7.8804300 & 110 & -7.8804300 \\
        150 & -7.8861508 & 82 & -7.8804300 & 112 & -7.8804300\\
        200 & -7.8806642 & 84 & -7.8804300 & & \\
        250 & -7.8804379 & 86 & -7.8804300 & & \\
        300 & -7.8804302 & & & & \\
        350 & -7.8804298 & & & & \\
    \end{tabular}
    \end{ruledtabular}
\end{table}

\begin{figure}
    \centering
    \includegraphics[width=\linewidth]{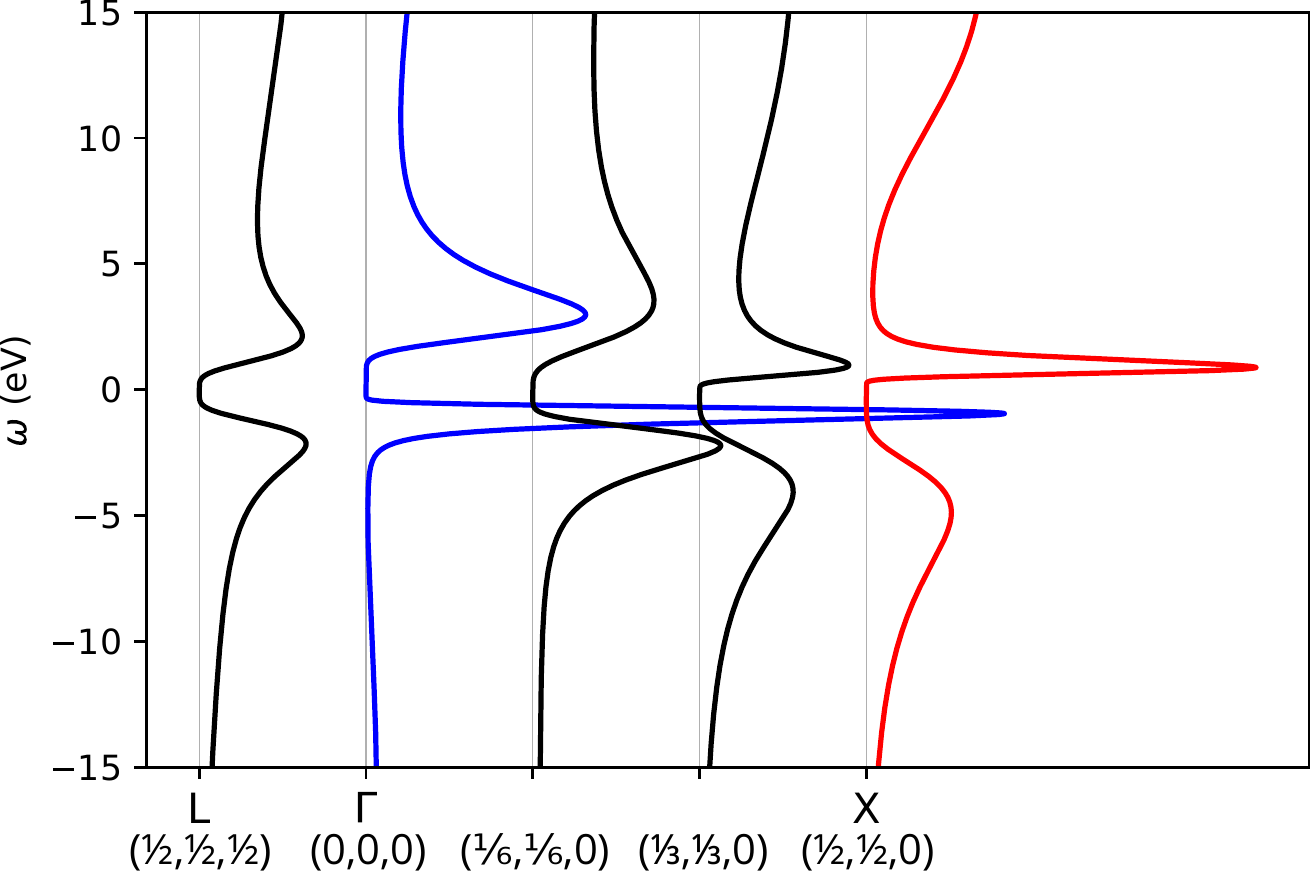}
    \caption{Momentum resolved spectral function of silicon at $\beta=\SI{1000}{\per\hartree}$, calculated from converged $GW$ Green's functions. Blue and red lines highlight spectra at $\Gamma$ and X points, which are the locations of the indirect band gap.
    }
    \label{fig:silicon-bandstructure}
\end{figure}

The sparse sampling method thus allows us to obtain numerically precise values for physical quantities with much less computational cost.
Fig.~\ref{fig:silicon-bandstructure} shows the momentum-resolved spectral functions calculated from the converged \GW{} Green's function via analytical continuation to the real frequency axis using the maximum entropy (MaxEnt) method.
The indirect band gap of the silicon crystal between $\Gamma$ and X points is well captured, with an estimated value of 1.84~eV.

\section{Conclusions}

We have introduced a sparse sampling scheme for
imaginary-time Green's functions which greatly reduces the size of imaginary time and frequency grids in realistic calculations.
We have developed general procedures to generate imaginary time and frequency sampling points from basis representations such as Chebyshev and IR, with efficient transformations between time and frequency domain.
We have applied the procedure to low-order diagrammatic methods and demonstrated the numerical performance in realistic molecules and solids.
With no more than a few hundred sampling points, the sparse sampling scheme can accurately capture physics in realistic systems with large energy scales and at low temperatures.

The sparse sampling reveals the full power of compact basis representations of the Green's function as the compactness is maintained in all stages of a calculation.
The number of sampling points necessary for the required precision scales slowly as temperature is lowered and energy scales are increased.
Our work therefore allows finite-temperature many-body methods to access physics of correlated electrons at larger energy scale and lower temperature.

Potential applications of the sparse sampling scheme include state-of-the-art methods for strongly correlated materials such as \GW+DMFT~\cite{Sun2002,Biermann02,Tomczak_2012,Tomczak2017,Werner2016} and the self-energy embedding theory (SEET)~\cite{Lan:2017hk}.
Quantitative estimates of superconducting temperature $T_\mathrm{c}$ based on the Migdal-Eliashberg theory in \textit{ab initio} calculations may benefit from the sparse sampling scheme so that the constant density of states approximation is no longer necessary~\cite{sano2016eliashberg}.

In addition, two-particle quantities play a key role in many diagrammatic calculations such as computing lattice susceptibility in DMFT~\cite{Jarrell92} as well as diagrammatic extensions of DMFT~\cite{Karsten2008, Rubtsov:2008cs}.
Two-particle quantities are difficult to handle even in effective model calculations due to the multiple indices for frequencies and spin/orbital degrees of freedom.
The IR basis has recently been extended to two-particle quantities~\cite{Shinaoka:2018cg}.
The application of the sparse sampling scheme to two-particle quantities is an interesting topic for future research.

\begin{acknowledgments}
    HS is grateful to R. Arita, D. Geffroy, J. Kune\v{s}, T. Koretsune, Y. Nagai, T. Nomoto, J. Otsuki and K. Yoshimi for fruitful discussions.
    HS was supported
    by JSPS KAKENHI Grant Nos. 16H01064 (J-Physics), 18H04301 (J-Physics), 18H01158, 16K17735.
    JL, MW, CNY, and EG are supported by NSF DMR 1606348.
    This research partially used resources of the National
    Energy Research Scientific Computing Center, a DOE Office of Science User Facility supported by the Office of Science of the U.S. Department of Energy under Contract No. DE-AC02-05CH11231.
\end{acknowledgments}

\appendix

\section{Compact orthogonal representation of Green's function}\label{appendix:compact_basis}

\subsection{Chebyshev basis}
The Chebyshev polynomials of the first kind $T_l(x)$ form an orthogonal system in the interval $[-1, 1]$,  which can be mapped in to the interval $[0, \beta]$ via
\begin{equation}
    x(\tau) = \frac{2\tau}{\beta}-1,\quad \tau(x) = \frac{\beta(x+1)}{2}
    \label{eq:tau-mapping}
\end{equation}
such that $F^\alpha_l(\tau) = T_l[x(\tau)]$.
We use the notation $T_l(\tau)$ to represent the order $l$ Chebyshev polynomial mapped onto the interval $[0,\beta]$.

Approximating an analytical function with the first $N$ Chebyshev polynomials is convenient due to the discrete orthogonality on the roots of the $(N+1)$-th Chebyshev polynomial
\begin{equation}
    \frac{1}{N}\sum_{k=0}^{N-1} T_i(x_k) T_j(x_k) = \frac{1+\delta_{i,0}}{2}\delta_{ij}
    \label{eq:cheb-ortho}
\end{equation}
where $x_k$ are the roots of $T_N(x)$. The Chebyshev coefficients are therefore well approximated by Clenshaw--Curtis quadrature:
\begin{equation}
    G^\alpha_l = \frac{2}{N(1+\delta_{0,l})}\sum_{k=1}^{N-1} G^\alpha(\tau_k) T_l(\tau_k) + \mathcal O(2^{-N})
    \label{eq:cheb-fit-root}
\end{equation}
where $\tau_k = \tau(x_k)$ defined in (\ref{eq:tau-mapping}).

With the Chebyshev coefficients $G^\alpha_l$, one can perform fast interpolation of $G^\alpha(\tau)$ at any $\tau\in[0,\beta]$ using recursion relations. Fourier transforms of the basis function $\hat{T}^\alpha_{l}(i\omega^\alpha_n)$ can also be computed, see Ref.~\onlinecite{gull2018chebyshev}.

\subsection{IR basis}
The IR basis introduced in Refs.~\onlinecite{shinaoka2017compressing} is designed to better capture properties of Green's functions in physical systems rather than arbitrary analytic functions.
The IR basis has been applied to numerical analytic continuation~\cite{Otsuki:2017er} and DMFT calculations~\cite{Nagai:2019dea}.
This section provides a brief description of the IR basis following the notation used in Ref.~\onlinecite{chikano2019irbasis}.
The IR basis originates from the Lehmann
representation of the single-particle Green’s function 
\begin{equation}
  G^{\alpha}(\tau) = -\int^{\wmax}_{-\wmax}d\omega
  K^{\alpha}(\tau,\omega)\rho^\alpha(\omega),
  \label{eq:spectral}
\end{equation}
where the spectrum $\rho^{\alpha}(\omega)$ is bounded in the interval $[-\wmax, \wmax]$ ($\wmax$ is a cutoff frequency).
The kernel $K^{\alpha}(\tau,\omega)$ reads
\begin{align}
  K^\alpha(\tau, \omega)
  &\equiv \omega^{\delta_{\alpha, \mathrm{B}}}
    \frac{e^{-\tau\omega}}{1 \pm e^{-\beta
    \omega}}
    \label{eq:K}
\end{align}
for $\tau \in [0, \beta]$,
where the $+$ and $-$ signs are used for fermions and bosons, respectively.
The extra $\omega$ factor for bosons in Eq. (\ref{eq:K})
is introduced in order to avoid a singularity of the kernel at $\omega=0$.

For a fixed value of $\beta$ and $\wmax$,
the IR basis functions are defined through the singular value decomposition (SVD)
\begin{align}
K^\alpha(\tau, \omega) = \sum^{\infty}_{l=0} S^\alpha_l U^{\alpha}_l(\tau) V^{\alpha}_l(\omega)
\label{eq:decomp-UV}
\end{align}
where one observes an exponential decay of the singular values $S_l^\alpha$ ($>0$) with increasing $l$. $U_l(\tau)$ and $V_l(\omega)$ form an orthonormal system for $\tau \in [0, \beta]$ and $y\in[-\wmax,\wmax]$, respectively.

A Green's function can be expanded as
\begin{align}
  G^{\alpha}(\tau) &= \sum_{l=0}^\infty G_l^{\alpha} U_l^{\alpha}(\tau) \label{eq:IR-decomp},\\
  G_l^\alpha &= - S_l^\alpha \rho_l^\alpha,\label{eq:Gl_rhol}
\end{align}
where
\begin{align}
   \rho_l^\alpha &\equiv \int_{-\wmax}^\wmax d\omega \rho^\alpha(\omega)   V_l^\alpha(\omega).\label{eq:rhol}
\end{align}
If $|\rho^\alpha_l|$ does not grow,
the exponential decay of $S_l^{\alpha}$ ensures exponential convergence of $G_l^\alpha$.
The accuracy of the expansion can be controlled by applying a cut-off on the singular values.

In calculations of realistic systems, one can set $\wmax$ large enough to capture the expected spectral width.
The basis functions change their shapes through the change of the dimensionless quantity $\Lambda = \beta\wmax$ as temperature is lowered.
This leads to a logarithmic growth of the basis size with respect to $\beta$.

The IR basis $\hat{U}^\alpha_l(i\omega^\alpha_n)$ does not compactly describe a constant shift in Matsubara frequency which corresponds to an unbounded spectrum.
Thus, any constant term must be subtracted beforehand particularly when expanding the self-energy.
Also, $U^\mathrm{B}_l(\tau)$ does not describe a constant shift in imaginary time, corresponding to a zero-energy mode.
Such terms must be treated separately as well~\cite{chikano2018performance}.

The dimensionless form of the IR basis is defined as
\begin{align}
    U^\alpha_l(\tau) &= \sqrt{\frac{2}{\beta}} u^\alpha_l(x(\tau)),\\
    \hat{U}_l^\alpha(i\omega^\alpha_n) &= \sqrt{\beta}u^\alpha_{ln} \nonumber\\
    &= \sqrt{\frac{\beta}{2}} \int_{-1}^{1} d x e^{\mathi \pi \{n+(1/2)\deltaF\}(x+1)} u^\alpha_l(x),
\end{align}
where $u_l^\alpha(x)$ form an orthonormalized basis for $x \in [-1, 1]$.

\section{Sampling points and condition numbers}%
\label{appendix:sample-detail}

\subsection{Matsubara sampling points for Chebyshev}

In the Chebyshev representation, we follow the same heuristics as in the $\tau$ sampling by finding or approximating zeros of the next basis function $\hat{T}^\alpha_{N}(i\omega^\alpha_n)$.
The following properties can be shown using the recursion relation developed in Ref.~\onlinecite{gull2018chebyshev}.
\begin{itemize}
    \item $\hat{T}^\alpha_{l}(i\omega^\alpha_n)$ can be written as a polynomial $I^\alpha_l(z_n) = \hat{T}^\alpha_{l}(z_n^{-1})$ with respect to the inverse frequency $z_n = 1/\omega^\alpha_n$.
    \item In the fermionic case, $I^{\mathrm{F}}_l(z)$ has order $l+1$. It has $l$ roots away from $z=0$
    when $l$ is even, and $l-1$ when $l$ is odd.
    \item In the bosonic case, $I^{\mathrm{B}}_l(z)$ has order $l$. It has $l-1$ roots away from $z=0$ when $l$ is odd, and $l-2$ when $l$ is even except for $l=0$, where there is no root.
\end{itemize}
Since $z_n \to 0$ implies $\omega^\alpha_n$ going to infinity, we cannot use $z = 0$ to define our sampling points. In order to take full advantage of the root structure of $I^\alpha_l(z)$ while being able to have at least $N$ sampling points, we put the following restrictions on the basis size $N$, based on whether fermionic or bosonic statistics is concerned.

In the fermionic case, we require that basis size $N$ to be even, such that $I^{\mathrm{F}}_{N}(z)$ has exactly $N$ roots $\{z_k\}$ away from $z=0$.
In general, $z_k^{-1}$ do not coincide exactly with Matsubara frequencies. We therefore approximate them by choosing the sampling points $\wbar^{\mathrm{F}}_k$ to be the closest Matsubara frequency to $z_k^{-1}$.

In the bosonic case, even if we require $N$ to be odd, there still are only $N-1$ roots away from $z=0$ from $I^{\mathrm{B}}_N(z)$.
The zero bosonic frequency $i\omega^{\mathrm{B}}_n = 0$, which corresponds to a constant shift in $\tau$, serves as a natural complement.
The sampling points $\wbar^{\mathrm{B}}_k$ is therefore composed of $N-1$ non-zero frequencies from approximating roots of $I^{\mathrm{B}}_l(z)$ and one zero-frequency point.

Following this algorithm, we are able to get exactly $N$ sampling points for either fermionic or bosonic statistics.

\subsection{Condition numbers of transformation matrices}

\begin{figure}
    \centering
    \includegraphics[width=\linewidth]{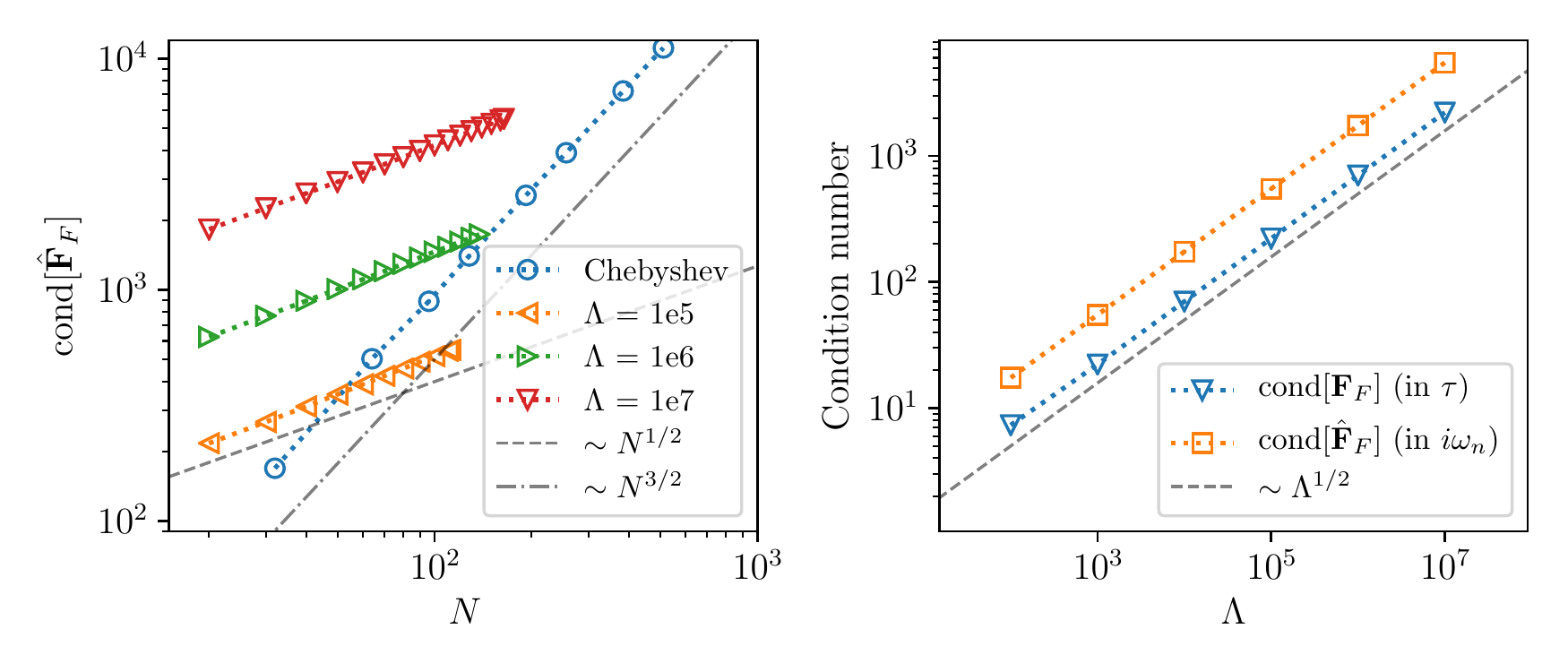}
    \caption{Condition number of the transformation matrices.
    Left panel shows the condition number of frequency transformation matrices  as a function of basis size $N$, for both Chebyshev and IR.
    Right panel shows the condition number of both $\tau$ and $i\omega_n$ transformation matrices with respect to $\Lambda$ for the fermionic IR basis.
    $N$ is chosen to be the maximum number of coefficients with the same cutoff in singular, provided in the \texttt{irbasis} library.
    }
    \label{fig:cond_number}
\end{figure}

Every time  Eqs.~(\ref{eq:back-trans-tau}) and (\ref{eq:back-trans-iw}) are evaluated, numerical errors, such as round-off error in floating point operations, may be amplified due to the (pseudo-)inversion process.
This error amplification can be quantified by the condition number of the transformation matrices $\mathbf{F}_\alpha$ and $\hat{\mathbf{F}}_\alpha$, defined as the product of the 2-norms of the matrix and its inverse.
In Fig.~\ref{fig:cond_number} we show the behaviors of the condition numbers as a function of the basis size $N$ for Chebyshev and IR (left panel), and as a function of $\Lambda$ for IR basis (right panel).
We can see that up to a significant number of basis functions, the condition numbers are $< 10^4$, which indicates well-conditioned inversion problems.
Since the values of $N$ and $\Lambda$ shown in Fig.~\ref{fig:cond_number} cover most values used in the calculations for this paper, the sparse sampling scheme guarantees stable numerical routines to get accurate results.

We observe that the condition number scales as $\calO(N^{3/2})$ for Chebyshev, and $\calO(N^{1/2})$ for IR. It also shows $\calO(\Lambda^{1/2})$ scaling for IR.
Although the origin of the condition number still requires further analysis, we believe that the algorithms we propose in this paper yield stable and predictable numerical procedures.

\section{Technical details of self-consistent calculations using the sparse sampling method}
\label{appendix:self-consistent-detail}

\subsection{Transforming between fermionic and bosonic statistics}

Besides the transformation matrices defined in Eqs.~(\ref{eq:trans-mat-tau}) and (\ref{eq:trans-mat-iw}), two additional matrices may be precomputed to allow fast switching between fermionic and bosonic representations in \GW{}
\begin{align}
    [\mathbf{F}^{\mathrm{F}\to \mathrm{B}}]_{kl} &= F^{\mathrm{F}}_l(\taubar^{\mathrm{B}}_k)\\
    [\mathbf{F}^{\mathrm{B}\to \mathrm{F}}]_{kl} &= F^{\mathrm{B}}_l(\taubar^{\mathrm{F}}_k).
\end{align}
Note that the inverse transform of those matrices are not well defined in general, since sampling points generated for one type of statistics usually do not serve as good sampling points for the other.

\subsection{Evaluation of total energy and density matrix}

In the total energy evaluation (\ref{eq:etotal}), the frequency summation term can be rewritten using an auxiliary scalar quantity $S$ such that
\begin{align}
    \frac{1}{2\beta}\sum_{n}\tr{[\hat{\tilde{\Sigma}}(i\omega^{\mathrm{F}}_n) \hat{G}(i\omega^{\mathrm{F}}_n)]} &= \frac{1}{2\beta}\sum_{n}\hat{S}(i\omega^{\mathrm{F}}_n) \nonumber\\
    &= \frac{1}{2}S(0^-) = -\frac{1}{2}S(\beta)
\end{align}
where
\begin{equation}
    \hat{S}(i\omega^{\mathrm{F}}_n) = \tr{[\hat{\tilde{\Sigma}}(i\omega^{\mathrm{F}}_n) \hat{G}(i\omega^{\mathrm{F}}_n)]}.
\end{equation}
We first evaluate $\hat{S}(i\wbar^{\mathrm{F}}_k)$ on the frequency sampling points $i\wbar^{\mathrm{F}}_k$, which is then transformed to the basis representation $S^{\mathrm{F}}_l$. The value of $S(\beta)$ is now a straightforward basis expansion at $\tau=\beta$
\begin{equation}
    S(\beta) = \sum_{l=0}^{N-1} S^{\mathrm{F}}_l F^{\mathrm{F}}_l(\beta).
\end{equation}
For the Chebyshev basis we simply have $T_l(\beta) = 1$.
In the case of IR basis, it is desirable to also tabulate the values $U^{\mathrm{F}}_l(\beta)$ along with the transformation matrices for efficient evaluations of relevant quantities.

Similarly, the calculation of the density matrix $\rho$ is a straightforward evaluation at $\tau=\beta$ from $G^{\mathrm{F}}_l$.
In calculations where the number of electrons is fixed, the chemical potential $\mu$ needs to be adjusted in each self-consistent iteration through a root finding procedure to conserve particle number~\cite{phillips2014communication}.
This step involves repeated density evaluations and solutions of the Dyson equation using the frequency-dependent self-energy, and sometimes becomes the bottleneck of the calculation.
The sparse sampling scheme massively reduces the number of frequency points needed in this process, which leads to a significant speedup over traditional approaches.

\bibliography{ref}

\begin{thebibliography}{68}%
\makeatletter
\providecommand \@ifxundefined [1]{%
 \@ifx{#1\undefined}
}%
\providecommand \@ifnum [1]{%
 \ifnum #1\expandafter \@firstoftwo
 \else \expandafter \@secondoftwo
 \fi
}%
\providecommand \@ifx [1]{%
 \ifx #1\expandafter \@firstoftwo
 \else \expandafter \@secondoftwo
 \fi
}%
\providecommand \natexlab [1]{#1}%
\providecommand \enquote  [1]{``#1''}%
\providecommand \bibnamefont  [1]{#1}%
\providecommand \bibfnamefont [1]{#1}%
\providecommand \citenamefont [1]{#1}%
\providecommand \href@noop [0]{\@secondoftwo}%
\providecommand \href [0]{\begingroup \@sanitize@url \@href}%
\providecommand \@href[1]{\@@startlink{#1}\@@href}%
\providecommand \@@href[1]{\endgroup#1\@@endlink}%
\providecommand \@sanitize@url [0]{\catcode `\\12\catcode `\$12\catcode
  `\&12\catcode `\#12\catcode `\^12\catcode `\_12\catcode `\%12\relax}%
\providecommand \@@startlink[1]{}%
\providecommand \@@endlink[0]{}%
\providecommand \url  [0]{\begingroup\@sanitize@url \@url }%
\providecommand \@url [1]{\endgroup\@href {#1}{\urlprefix }}%
\providecommand \urlprefix  [0]{URL }%
\providecommand \Eprint [0]{\href }%
\providecommand \doibase [0]{https://doi.org/}%
\providecommand \selectlanguage [0]{\@gobble}%
\providecommand \bibinfo  [0]{\@secondoftwo}%
\providecommand \bibfield  [0]{\@secondoftwo}%
\providecommand \translation [1]{[#1]}%
\providecommand \BibitemOpen [0]{}%
\providecommand \bibitemStop [0]{}%
\providecommand \bibitemNoStop [0]{.\EOS\space}%
\providecommand \EOS [0]{\spacefactor3000\relax}%
\providecommand \BibitemShut  [1]{\csname bibitem#1\endcsname}%
\let\auto@bib@innerbib\@empty
\bibitem [{\citenamefont {Matsubara}(1955)}]{Matsubara55}%
  \BibitemOpen
  \bibfield  {author} {\bibinfo {author} {\bibfnamefont {T.}~\bibnamefont
  {Matsubara}},\ }\bibfield  {title} {\bibinfo {title} {{A New Approach to
  Quantum-Statistical Mechanics}},\ }\href {https://doi.org/10.1143/PTP.14.351}
  {\bibfield  {journal} {\bibinfo  {journal} {Progress of Theoretical Physics}\
  }\textbf {\bibinfo {volume} {14}},\ \bibinfo {pages} {351} (\bibinfo {year}
  {1955})}\BibitemShut {NoStop}%
\bibitem [{\citenamefont {Abrikosov}\ \emph {et~al.}(1975)\citenamefont
  {Abrikosov}, \citenamefont {Gorkov},\ and\ \citenamefont
  {Dzyaloshinski}}]{abrikosov1975methods}%
  \BibitemOpen
  \bibfield  {author} {\bibinfo {author} {\bibfnamefont {A.~A.}\ \bibnamefont
  {Abrikosov}}, \bibinfo {author} {\bibfnamefont {L.~P.}\ \bibnamefont
  {Gorkov}},\ and\ \bibinfo {author} {\bibfnamefont {I.~E.}\ \bibnamefont
  {Dzyaloshinski}},\ }\href@noop {} {\emph {\bibinfo {title} {{Methods of
  Quantum Field Theory in Statistical Physics}}}}\ (\bibinfo  {publisher}
  {Dover Publications, New York},\ \bibinfo {year} {1975})\BibitemShut
  {NoStop}%
\bibitem [{\citenamefont {Negele}\ and\ \citenamefont
  {Orland}(1998)}]{negele1998quantum}%
  \BibitemOpen
  \bibfield  {author} {\bibinfo {author} {\bibfnamefont {J.~W.}\ \bibnamefont
  {Negele}}\ and\ \bibinfo {author} {\bibfnamefont {H.}~\bibnamefont
  {Orland}},\ }\href@noop {} {\emph {\bibinfo {title} {{Quantum many-particle
  systems (Advanced Book classics)}}}}\ (\bibinfo  {publisher} {Westview
  Press},\ \bibinfo {year} {1998})\BibitemShut {NoStop}%
\bibitem [{\citenamefont {Luttinger}\ and\ \citenamefont
  {Ward}(1960)}]{luttingerward1960}%
  \BibitemOpen
  \bibfield  {author} {\bibinfo {author} {\bibfnamefont {J.~M.}\ \bibnamefont
  {Luttinger}}\ and\ \bibinfo {author} {\bibfnamefont {J.~C.}\ \bibnamefont
  {Ward}},\ }\bibfield  {title} {\bibinfo {title} {{Ground-State Energy of a
  Many-Fermion System. II}},\ }\href {https://doi.org/10.1103/PhysRev.118.1417}
  {\bibfield  {journal} {\bibinfo  {journal} {Physical Review}\ }\textbf
  {\bibinfo {volume} {118}},\ \bibinfo {pages} {1417} (\bibinfo {year}
  {1960})}\BibitemShut {NoStop}%
\bibitem [{\citenamefont {Scalapino}\ and\ \citenamefont
  {Sugar}(1981)}]{blankenbecler1981monte}%
  \BibitemOpen
  \bibfield  {author} {\bibinfo {author} {\bibfnamefont {D.~J.}\ \bibnamefont
  {Scalapino}}\ and\ \bibinfo {author} {\bibfnamefont {R.~L.}\ \bibnamefont
  {Sugar}},\ }\bibfield  {title} {\bibinfo {title} {{Monte Carlo calculations
  of coupled boson-fermion systems. II}},\ }\href
  {https://doi.org/10.1103/PhysRevB.24.4295} {\bibfield  {journal} {\bibinfo
  {journal} {Physical Review B}\ }\textbf {\bibinfo {volume} {24}},\ \bibinfo
  {pages} {4295} (\bibinfo {year} {1981})}\BibitemShut {NoStop}%
\bibitem [{\citenamefont {Georges}\ \emph {et~al.}(1996)\citenamefont
  {Georges}, \citenamefont {Kotliar}, \citenamefont {Krauth},\ and\
  \citenamefont {Rozenberg}}]{georges1996dynamical}%
  \BibitemOpen
  \bibfield  {author} {\bibinfo {author} {\bibfnamefont {A.}~\bibnamefont
  {Georges}}, \bibinfo {author} {\bibfnamefont {G.}~\bibnamefont {Kotliar}},
  \bibinfo {author} {\bibfnamefont {W.}~\bibnamefont {Krauth}},\ and\ \bibinfo
  {author} {\bibfnamefont {M.~J.}\ \bibnamefont {Rozenberg}},\ }\bibfield
  {title} {\bibinfo {title} {{Dynamical mean-field theory of strongly
  correlated fermion systems and the limit of infinite dimensions}},\ }\href
  {https://doi.org/10.1103/RevModPhys.68.13} {\bibfield  {journal} {\bibinfo
  {journal} {Reviews of Modern Physics}\ }\textbf {\bibinfo {volume} {68}},\
  \bibinfo {pages} {13} (\bibinfo {year} {1996})}\BibitemShut {NoStop}%
\bibitem [{\citenamefont {Maier}\ \emph {et~al.}(2005)\citenamefont {Maier},
  \citenamefont {Jarrell}, \citenamefont {Pruschke},\ and\ \citenamefont
  {Hettler}}]{Maier05}%
  \BibitemOpen
  \bibfield  {author} {\bibinfo {author} {\bibfnamefont {T.}~\bibnamefont
  {Maier}}, \bibinfo {author} {\bibfnamefont {M.}~\bibnamefont {Jarrell}},
  \bibinfo {author} {\bibfnamefont {T.}~\bibnamefont {Pruschke}},\ and\
  \bibinfo {author} {\bibfnamefont {M.~H.}\ \bibnamefont {Hettler}},\
  }\bibfield  {title} {\bibinfo {title} {{Quantum cluster theories}},\ }\href
  {https://doi.org/10.1103/RevModPhys.77.1027} {\bibfield  {journal} {\bibinfo
  {journal} {Reviews of Modern Physics}\ }\textbf {\bibinfo {volume} {77}},\
  \bibinfo {pages} {1027} (\bibinfo {year} {2005})}\BibitemShut {NoStop}%
\bibitem [{\citenamefont {Kotliar}\ \emph {et~al.}(2006)\citenamefont
  {Kotliar}, \citenamefont {Savrasov}, \citenamefont {Haule}, \citenamefont
  {Oudovenko}, \citenamefont {Parcollet},\ and\ \citenamefont
  {Marianetti}}]{Kotliar06}%
  \BibitemOpen
  \bibfield  {author} {\bibinfo {author} {\bibfnamefont {G.}~\bibnamefont
  {Kotliar}}, \bibinfo {author} {\bibfnamefont {S.~Y.}\ \bibnamefont
  {Savrasov}}, \bibinfo {author} {\bibfnamefont {K.}~\bibnamefont {Haule}},
  \bibinfo {author} {\bibfnamefont {V.~S.}\ \bibnamefont {Oudovenko}}, \bibinfo
  {author} {\bibfnamefont {O.}~\bibnamefont {Parcollet}},\ and\ \bibinfo
  {author} {\bibfnamefont {C.~A.}\ \bibnamefont {Marianetti}},\ }\bibfield
  {title} {\bibinfo {title} {{Electronic structure calculations with dynamical
  mean-field theory}},\ }\href {https://doi.org/10.1103/RevModPhys.78.865}
  {\bibfield  {journal} {\bibinfo  {journal} {Reviews of Modern Physics}\
  }\textbf {\bibinfo {volume} {78}},\ \bibinfo {pages} {865} (\bibinfo {year}
  {2006})}\BibitemShut {NoStop}%
\bibitem [{\citenamefont {Held}\ \emph {et~al.}(2006)\citenamefont {Held},
  \citenamefont {Nekrasov}, \citenamefont {Keller}, \citenamefont {Eyert},
  \citenamefont {Bl{\"{u}}mer}, \citenamefont {McMahan}, \citenamefont
  {Scalettar}, \citenamefont {Pruschke}, \citenamefont {Anisimov},\ and\
  \citenamefont {Vollhardt}}]{Held06}%
  \BibitemOpen
  \bibfield  {author} {\bibinfo {author} {\bibfnamefont {K.}~\bibnamefont
  {Held}}, \bibinfo {author} {\bibfnamefont {I.~A.}\ \bibnamefont {Nekrasov}},
  \bibinfo {author} {\bibfnamefont {G.}~\bibnamefont {Keller}}, \bibinfo
  {author} {\bibfnamefont {V.}~\bibnamefont {Eyert}}, \bibinfo {author}
  {\bibfnamefont {N.}~\bibnamefont {Bl{\"{u}}mer}}, \bibinfo {author}
  {\bibfnamefont {A.~K.}\ \bibnamefont {McMahan}}, \bibinfo {author}
  {\bibfnamefont {R.~T.}\ \bibnamefont {Scalettar}}, \bibinfo {author}
  {\bibfnamefont {T.}~\bibnamefont {Pruschke}}, \bibinfo {author}
  {\bibfnamefont {V.~I.}\ \bibnamefont {Anisimov}},\ and\ \bibinfo {author}
  {\bibfnamefont {D.}~\bibnamefont {Vollhardt}},\ }\bibfield  {title} {\bibinfo
  {title} {{Realistic investigations of correlated electron systems with LDA +
  DMFT}},\ }\href {https://doi.org/10.1002/pssb.200642053} {\bibfield
  {journal} {\bibinfo  {journal} {physica status solidi (b)}\ }\textbf
  {\bibinfo {volume} {243}},\ \bibinfo {pages} {2599} (\bibinfo {year}
  {2006})}\BibitemShut {NoStop}%
\bibitem [{\citenamefont {Hafermann}\ \emph {et~al.}(2008)\citenamefont
  {Hafermann}, \citenamefont {Brener}, \citenamefont {Rubtsov}, \citenamefont
  {Katsnelson},\ and\ \citenamefont {Lichtenstein}}]{Rubtsov:2008cs}%
  \BibitemOpen
  \bibfield  {author} {\bibinfo {author} {\bibfnamefont {H.}~\bibnamefont
  {Hafermann}}, \bibinfo {author} {\bibfnamefont {S.}~\bibnamefont {Brener}},
  \bibinfo {author} {\bibfnamefont {A.~N.}\ \bibnamefont {Rubtsov}}, \bibinfo
  {author} {\bibfnamefont {M.~I.}\ \bibnamefont {Katsnelson}},\ and\ \bibinfo
  {author} {\bibfnamefont {A.~I.}\ \bibnamefont {Lichtenstein}},\ }\bibfield
  {title} {\bibinfo {title} {{Cluster dual fermion approach to nonlocal
  correlations}},\ }\href {https://doi.org/10.1134/s0021364007220134}
  {\bibfield  {journal} {\bibinfo  {journal} {JETP Letters}\ }\textbf {\bibinfo
  {volume} {86}},\ \bibinfo {pages} {677} (\bibinfo {year} {2008})}\BibitemShut
  {NoStop}%
\bibitem [{\citenamefont {Toschi}\ \emph {et~al.}(2007)\citenamefont {Toschi},
  \citenamefont {Katanin},\ and\ \citenamefont {Held}}]{Toschi07}%
  \BibitemOpen
  \bibfield  {author} {\bibinfo {author} {\bibfnamefont {A.}~\bibnamefont
  {Toschi}}, \bibinfo {author} {\bibfnamefont {A.~A.}\ \bibnamefont
  {Katanin}},\ and\ \bibinfo {author} {\bibfnamefont {K.}~\bibnamefont
  {Held}},\ }\bibfield  {title} {\bibinfo {title} {{Dynamical vertex
  approximation: A step beyond dynamical mean-field theory}},\ }\href
  {https://doi.org/10.1103/PhysRevB.75.045118} {\bibfield  {journal} {\bibinfo
  {journal} {Physical Review B}\ }\textbf {\bibinfo {volume} {75}},\ \bibinfo
  {pages} {045118} (\bibinfo {year} {2007})}\BibitemShut {NoStop}%
\bibitem [{\citenamefont {Rohringer}\ \emph {et~al.}(2018)\citenamefont
  {Rohringer}, \citenamefont {Hafermann}, \citenamefont {Toschi}, \citenamefont
  {Katanin}, \citenamefont {Antipov}, \citenamefont {Katsnelson}, \citenamefont
  {Lichtenstein}, \citenamefont {Rubtsov},\ and\ \citenamefont
  {Held}}]{Rohringer18}%
  \BibitemOpen
  \bibfield  {author} {\bibinfo {author} {\bibfnamefont {G.}~\bibnamefont
  {Rohringer}}, \bibinfo {author} {\bibfnamefont {H.}~\bibnamefont
  {Hafermann}}, \bibinfo {author} {\bibfnamefont {A.}~\bibnamefont {Toschi}},
  \bibinfo {author} {\bibfnamefont {A.~A.}\ \bibnamefont {Katanin}}, \bibinfo
  {author} {\bibfnamefont {A.~E.}\ \bibnamefont {Antipov}}, \bibinfo {author}
  {\bibfnamefont {M.~I.}\ \bibnamefont {Katsnelson}}, \bibinfo {author}
  {\bibfnamefont {A.~I.}\ \bibnamefont {Lichtenstein}}, \bibinfo {author}
  {\bibfnamefont {A.~N.}\ \bibnamefont {Rubtsov}},\ and\ \bibinfo {author}
  {\bibfnamefont {K.}~\bibnamefont {Held}},\ }\bibfield  {title} {\bibinfo
  {title} {{Diagrammatic routes to nonlocal correlations beyond dynamical mean
  field theory}},\ }\href {https://doi.org/10.1103/RevModPhys.90.025003}
  {\bibfield  {journal} {\bibinfo  {journal} {Reviews of Modern Physics}\
  }\textbf {\bibinfo {volume} {90}},\ \bibinfo {pages} {25003} (\bibinfo {year}
  {2018})}\BibitemShut {NoStop}%
\bibitem [{\citenamefont {Prokof'ev}\ and\ \citenamefont
  {Svistunov}(1998)}]{prokof1998polaron}%
  \BibitemOpen
  \bibfield  {author} {\bibinfo {author} {\bibfnamefont {N.~V.}\ \bibnamefont
  {Prokof'ev}}\ and\ \bibinfo {author} {\bibfnamefont {B.~V.}\ \bibnamefont
  {Svistunov}},\ }\bibfield  {title} {\bibinfo {title} {{Polaron problem by
  diagrammatic quantum monte carlo}},\ }\href
  {https://doi.org/10.1103/PhysRevLett.81.2514} {\bibfield  {journal} {\bibinfo
   {journal} {Physical Review Letters}\ }\textbf {\bibinfo {volume} {81}},\
  \bibinfo {pages} {2514} (\bibinfo {year} {1998})}\BibitemShut {NoStop}%
\bibitem [{\citenamefont {Gull}\ \emph {et~al.}(2011)\citenamefont {Gull},
  \citenamefont {Millis}, \citenamefont {Lichtenstein}, \citenamefont
  {Rubtsov}, \citenamefont {Troyer},\ and\ \citenamefont
  {Werner}}]{gull2011continuous}%
  \BibitemOpen
  \bibfield  {author} {\bibinfo {author} {\bibfnamefont {E.}~\bibnamefont
  {Gull}}, \bibinfo {author} {\bibfnamefont {A.~J.}\ \bibnamefont {Millis}},
  \bibinfo {author} {\bibfnamefont {A.~I.}\ \bibnamefont {Lichtenstein}},
  \bibinfo {author} {\bibfnamefont {A.~N.}\ \bibnamefont {Rubtsov}}, \bibinfo
  {author} {\bibfnamefont {M.}~\bibnamefont {Troyer}},\ and\ \bibinfo {author}
  {\bibfnamefont {P.}~\bibnamefont {Werner}},\ }\bibfield  {title} {\bibinfo
  {title} {{Continuous-time Monte Carlo methods for quantum impurity models}},\
  }\href {https://doi.org/10.1103/RevModPhys.83.349} {\bibfield  {journal}
  {\bibinfo  {journal} {Reviews of Modern Physics}\ }\textbf {\bibinfo {volume}
  {83}},\ \bibinfo {pages} {349} (\bibinfo {year} {2011})}\BibitemShut
  {NoStop}%
\bibitem [{\citenamefont {Hedin}(1965)}]{hedin1965gw}%
  \BibitemOpen
  \bibfield  {author} {\bibinfo {author} {\bibfnamefont {L.}~\bibnamefont
  {Hedin}},\ }\bibfield  {title} {\bibinfo {title} {{New Method for Calculating
  the One-Particle Green's Function with Application to the Electron-Gas
  Problem}},\ }\href {https://doi.org/10.1103/PhysRev.139.A796} {\bibfield
  {journal} {\bibinfo  {journal} {Physical Review}\ }\textbf {\bibinfo {volume}
  {139}},\ \bibinfo {pages} {A796} (\bibinfo {year} {1965})}\BibitemShut
  {NoStop}%
\bibitem [{\citenamefont {Aryasetiawan}\ and\ \citenamefont
  {Gunnarsson}(1998)}]{Aryasetiawan_1998}%
  \BibitemOpen
  \bibfield  {author} {\bibinfo {author} {\bibfnamefont {F.}~\bibnamefont
  {Aryasetiawan}}\ and\ \bibinfo {author} {\bibfnamefont {O.}~\bibnamefont
  {Gunnarsson}},\ }\bibfield  {title} {\bibinfo {title} {{The GW method}},\
  }\href {https://doi.org/10.1088/0034-4885/61/3/002} {\bibfield  {journal}
  {\bibinfo  {journal} {Reports on Progress in Physics}\ }\textbf {\bibinfo
  {volume} {61}},\ \bibinfo {pages} {237} (\bibinfo {year} {1998})}\BibitemShut
  {NoStop}%
\bibitem [{\citenamefont {Stan}\ \emph {et~al.}(2009)\citenamefont {Stan},
  \citenamefont {Dahlen},\ and\ \citenamefont {van Leeuwen}}]{Stan2009}%
  \BibitemOpen
  \bibfield  {author} {\bibinfo {author} {\bibfnamefont {A.}~\bibnamefont
  {Stan}}, \bibinfo {author} {\bibfnamefont {N.~E.}\ \bibnamefont {Dahlen}},\
  and\ \bibinfo {author} {\bibfnamefont {R.}~\bibnamefont {van Leeuwen}},\
  }\bibfield  {title} {\bibinfo {title} {{Levels of self-consistency in the GW
  approximation}},\ }\href {https://doi.org/10.1063/1.3089567} {\bibfield
  {journal} {\bibinfo  {journal} {The Journal of Chemical Physics}\ }\textbf
  {\bibinfo {volume} {130}},\ \bibinfo {pages} {114105} (\bibinfo {year}
  {2009})}\BibitemShut {NoStop}%
\bibitem [{\citenamefont {Kutepov}\ \emph {et~al.}(2009)\citenamefont
  {Kutepov}, \citenamefont {Savrasov},\ and\ \citenamefont
  {Kotliar}}]{Kutepov2009}%
  \BibitemOpen
  \bibfield  {author} {\bibinfo {author} {\bibfnamefont {A.}~\bibnamefont
  {Kutepov}}, \bibinfo {author} {\bibfnamefont {S.~Y.}\ \bibnamefont
  {Savrasov}},\ and\ \bibinfo {author} {\bibfnamefont {G.}~\bibnamefont
  {Kotliar}},\ }\bibfield  {title} {\bibinfo {title} {{Ground-state properties
  of simple elements from GW calculations}},\ }\href
  {https://doi.org/10.1103/PhysRevB.80.041103} {\bibfield  {journal} {\bibinfo
  {journal} {Physical Review B}\ }\textbf {\bibinfo {volume} {80}},\ \bibinfo
  {pages} {041103} (\bibinfo {year} {2009})}\BibitemShut {NoStop}%
\bibitem [{\citenamefont {{Van Setten}}\ \emph {et~al.}(2015)\citenamefont
  {{Van Setten}}, \citenamefont {Caruso}, \citenamefont {Sharifzadeh},
  \citenamefont {Ren}, \citenamefont {Scheffler}, \citenamefont {Liu},
  \citenamefont {Lischner}, \citenamefont {Lin}, \citenamefont {Deslippe},
  \citenamefont {Louie}, \citenamefont {Yang}, \citenamefont {Weigend},
  \citenamefont {Neaton}, \citenamefont {Evers},\ and\ \citenamefont
  {Rinke}}]{GW100}%
  \BibitemOpen
  \bibfield  {author} {\bibinfo {author} {\bibfnamefont {M.~J.}\ \bibnamefont
  {{Van Setten}}}, \bibinfo {author} {\bibfnamefont {F.}~\bibnamefont
  {Caruso}}, \bibinfo {author} {\bibfnamefont {S.}~\bibnamefont {Sharifzadeh}},
  \bibinfo {author} {\bibfnamefont {X.}~\bibnamefont {Ren}}, \bibinfo {author}
  {\bibfnamefont {M.}~\bibnamefont {Scheffler}}, \bibinfo {author}
  {\bibfnamefont {F.}~\bibnamefont {Liu}}, \bibinfo {author} {\bibfnamefont
  {J.}~\bibnamefont {Lischner}}, \bibinfo {author} {\bibfnamefont
  {L.}~\bibnamefont {Lin}}, \bibinfo {author} {\bibfnamefont {J.~R.}\
  \bibnamefont {Deslippe}}, \bibinfo {author} {\bibfnamefont {S.~G.}\
  \bibnamefont {Louie}}, \bibinfo {author} {\bibfnamefont {C.}~\bibnamefont
  {Yang}}, \bibinfo {author} {\bibfnamefont {F.}~\bibnamefont {Weigend}},
  \bibinfo {author} {\bibfnamefont {J.~B.}\ \bibnamefont {Neaton}}, \bibinfo
  {author} {\bibfnamefont {F.}~\bibnamefont {Evers}},\ and\ \bibinfo {author}
  {\bibfnamefont {P.}~\bibnamefont {Rinke}},\ }\bibfield  {title} {\bibinfo
  {title} {{GW100: Benchmarking $G_0 W_0$ for Molecular Systems}},\ }\href
  {https://doi.org/10.1021/acs.jctc.5b00453} {\bibfield  {journal} {\bibinfo
  {journal} {Journal of Chemical Theory and Computation}\ }\textbf {\bibinfo
  {volume} {11}},\ \bibinfo {pages} {5665} (\bibinfo {year}
  {2015})}\BibitemShut {NoStop}%
\bibitem [{\citenamefont {Maggio}\ \emph {et~al.}(2017)\citenamefont {Maggio},
  \citenamefont {Liu}, \citenamefont {van Setten},\ and\ \citenamefont
  {Kresse}}]{GW100pw}%
  \BibitemOpen
  \bibfield  {author} {\bibinfo {author} {\bibfnamefont {E.}~\bibnamefont
  {Maggio}}, \bibinfo {author} {\bibfnamefont {P.}~\bibnamefont {Liu}},
  \bibinfo {author} {\bibfnamefont {M.~J.}\ \bibnamefont {van Setten}},\ and\
  \bibinfo {author} {\bibfnamefont {G.}~\bibnamefont {Kresse}},\ }\bibfield
  {title} {\bibinfo {title} {{GW 100: A Plane Wave Perspective for Small
  Molecules}},\ }\href {https://doi.org/10.1021/acs.jctc.6b01150} {\bibfield
  {journal} {\bibinfo  {journal} {Journal of Chemical Theory and Computation}\
  }\textbf {\bibinfo {volume} {13}},\ \bibinfo {pages} {635} (\bibinfo {year}
  {2017})}\BibitemShut {NoStop}%
\bibitem [{\citenamefont {Grumet}\ \emph {et~al.}(2018)\citenamefont {Grumet},
  \citenamefont {Liu}, \citenamefont {Kaltak}, \citenamefont {Klime{\v{s}}},\
  and\ \citenamefont {Kresse}}]{Grumet2018}%
  \BibitemOpen
  \bibfield  {author} {\bibinfo {author} {\bibfnamefont {M.}~\bibnamefont
  {Grumet}}, \bibinfo {author} {\bibfnamefont {P.}~\bibnamefont {Liu}},
  \bibinfo {author} {\bibfnamefont {M.}~\bibnamefont {Kaltak}}, \bibinfo
  {author} {\bibfnamefont {J.}~\bibnamefont {Klime{\v{s}}}},\ and\ \bibinfo
  {author} {\bibfnamefont {G.}~\bibnamefont {Kresse}},\ }\bibfield  {title}
  {\bibinfo {title} {{Beyond the quasiparticle approximation: Fully
  self-consistent $GW$ calculations}},\ }\href
  {https://doi.org/10.1103/PhysRevB.98.155143} {\bibfield  {journal} {\bibinfo
  {journal} {Physical Review B}\ }\textbf {\bibinfo {volume} {98}},\ \bibinfo
  {pages} {155143} (\bibinfo {year} {2018})}\BibitemShut {NoStop}%
\bibitem [{\citenamefont {Kutepov}(2016)}]{Kutepov16}%
  \BibitemOpen
  \bibfield  {author} {\bibinfo {author} {\bibfnamefont {A.~L.}\ \bibnamefont
  {Kutepov}},\ }\bibfield  {title} {\bibinfo {title} {{Electronic structure of
  Na, K, Si, and LiF from self-consistent solution of Hedin's equations
  including vertex corrections}},\ }\href
  {https://doi.org/10.1103/PhysRevB.94.155101} {\bibfield  {journal} {\bibinfo
  {journal} {Physical Review B}\ }\textbf {\bibinfo {volume} {94}},\ \bibinfo
  {pages} {155101} (\bibinfo {year} {2016})}\BibitemShut {NoStop}%
\bibitem [{\citenamefont {Kutepov}(2017)}]{Kutepov17}%
  \BibitemOpen
  \bibfield  {author} {\bibinfo {author} {\bibfnamefont {A.~L.}\ \bibnamefont
  {Kutepov}},\ }\bibfield  {title} {\bibinfo {title} {{Self-consistent solution
  of Hedin's equations: Semiconductors and insulators}},\ }\href
  {https://doi.org/10.1103/PhysRevB.95.195120} {\bibfield  {journal} {\bibinfo
  {journal} {Physical Review B}\ }\textbf {\bibinfo {volume} {95}},\ \bibinfo
  {pages} {195120} (\bibinfo {year} {2017})}\BibitemShut {NoStop}%
\bibitem [{\citenamefont {Dahlen}\ and\ \citenamefont {{Van
  Leeuwen}}(2005)}]{dahlen2005gf2}%
  \BibitemOpen
  \bibfield  {author} {\bibinfo {author} {\bibfnamefont {N.~E.}\ \bibnamefont
  {Dahlen}}\ and\ \bibinfo {author} {\bibfnamefont {R.}~\bibnamefont {{Van
  Leeuwen}}},\ }\bibfield  {title} {\bibinfo {title} {{Self-consistent solution
  of the Dyson equation for atoms and molecules within a conserving
  approximation}},\ }\href {https://doi.org/10.1063/1.1884965} {\bibfield
  {journal} {\bibinfo  {journal} {Journal of Chemical Physics}\ }\textbf
  {\bibinfo {volume} {122}},\ \bibinfo {pages} {164102} (\bibinfo {year}
  {2005})}\BibitemShut {NoStop}%
\bibitem [{\citenamefont {Phillips}\ and\ \citenamefont
  {Zgid}(2014)}]{phillips2014communication}%
  \BibitemOpen
  \bibfield  {author} {\bibinfo {author} {\bibfnamefont {J.~J.}\ \bibnamefont
  {Phillips}}\ and\ \bibinfo {author} {\bibfnamefont {D.}~\bibnamefont
  {Zgid}},\ }\bibfield  {title} {\bibinfo {title} {{Communication: The
  description of strong correlation within self-consistent Green's function
  second-order perturbation theory}},\ }\href
  {https://doi.org/10.1063/1.4884951} {\bibfield  {journal} {\bibinfo
  {journal} {Journal of Chemical Physics}\ }\textbf {\bibinfo {volume} {140}},\
  \bibinfo {pages} {241101} (\bibinfo {year} {2014})}\BibitemShut {NoStop}%
\bibitem [{\citenamefont {Phillips}\ \emph {et~al.}(2015)\citenamefont
  {Phillips}, \citenamefont {Kananenka},\ and\ \citenamefont
  {Zgid}}]{phillips2015fractional}%
  \BibitemOpen
  \bibfield  {author} {\bibinfo {author} {\bibfnamefont {J.~J.}\ \bibnamefont
  {Phillips}}, \bibinfo {author} {\bibfnamefont {A.~A.}\ \bibnamefont
  {Kananenka}},\ and\ \bibinfo {author} {\bibfnamefont {D.}~\bibnamefont
  {Zgid}},\ }\bibfield  {title} {\bibinfo {title} {{Fractional charge and spin
  errors in self-consistent Green's function theory}},\ }\href
  {https://doi.org/10.1063/1.4921259} {\bibfield  {journal} {\bibinfo
  {journal} {Journal of Chemical Physics}\ }\textbf {\bibinfo {volume} {142}},\
  \bibinfo {pages} {194108} (\bibinfo {year} {2015})}\BibitemShut {NoStop}%
\bibitem [{\citenamefont {Kananenka}\ \emph
  {et~al.}(2016{\natexlab{a}})\citenamefont {Kananenka}, \citenamefont
  {Phillips},\ and\ \citenamefont {Zgid}}]{kananenka2016grid}%
  \BibitemOpen
  \bibfield  {author} {\bibinfo {author} {\bibfnamefont {A.~A.}\ \bibnamefont
  {Kananenka}}, \bibinfo {author} {\bibfnamefont {J.~J.}\ \bibnamefont
  {Phillips}},\ and\ \bibinfo {author} {\bibfnamefont {D.}~\bibnamefont
  {Zgid}},\ }\bibfield  {title} {\bibinfo {title} {{Efficient
  Temperature-Dependent Green's Functions Methods for Realistic Systems:
  Compact Grids for Orthogonal Polynomial Transforms}},\ }\href
  {https://doi.org/10.1021/acs.jctc.5b00884} {\bibfield  {journal} {\bibinfo
  {journal} {Journal of Chemical Theory and Computation}\ }\textbf {\bibinfo
  {volume} {12}},\ \bibinfo {pages} {564} (\bibinfo {year}
  {2016}{\natexlab{a}})}\BibitemShut {NoStop}%
\bibitem [{\citenamefont {Kananenka}\ \emph
  {et~al.}(2016{\natexlab{b}})\citenamefont {Kananenka}, \citenamefont
  {Welden}, \citenamefont {Lan}, \citenamefont {Gull},\ and\ \citenamefont
  {Zgid}}]{kananenka2016spline}%
  \BibitemOpen
  \bibfield  {author} {\bibinfo {author} {\bibfnamefont {A.~A.}\ \bibnamefont
  {Kananenka}}, \bibinfo {author} {\bibfnamefont {A.~R.}\ \bibnamefont
  {Welden}}, \bibinfo {author} {\bibfnamefont {T.~N.}\ \bibnamefont {Lan}},
  \bibinfo {author} {\bibfnamefont {E.}~\bibnamefont {Gull}},\ and\ \bibinfo
  {author} {\bibfnamefont {D.}~\bibnamefont {Zgid}},\ }\bibfield  {title}
  {\bibinfo {title} {{Efficient Temperature-Dependent Green's Function Methods
  for Realistic Systems: Using Cubic Spline Interpolation to Approximate
  Matsubara Green's Functions}},\ }\href
  {https://doi.org/10.1021/acs.jctc.6b00178} {\bibfield  {journal} {\bibinfo
  {journal} {Journal of Chemical Theory and Computation}\ }\textbf {\bibinfo
  {volume} {12}},\ \bibinfo {pages} {2250} (\bibinfo {year}
  {2016}{\natexlab{b}})}\BibitemShut {NoStop}%
\bibitem [{\citenamefont {Rusakov}\ and\ \citenamefont
  {Zgid}(2016)}]{rusakov2016periodic}%
  \BibitemOpen
  \bibfield  {author} {\bibinfo {author} {\bibfnamefont {A.~A.}\ \bibnamefont
  {Rusakov}}\ and\ \bibinfo {author} {\bibfnamefont {D.}~\bibnamefont {Zgid}},\
  }\bibfield  {title} {\bibinfo {title} {{Self-consistent second-order Green's
  function perturbation theory for periodic systems}},\ }\href
  {https://doi.org/10.1063/1.4940900} {\bibfield  {journal} {\bibinfo
  {journal} {Journal of Chemical Physics}\ }\textbf {\bibinfo {volume} {144}},\
  \bibinfo {pages} {54106} (\bibinfo {year} {2016})}\BibitemShut {NoStop}%
\bibitem [{\citenamefont {Welden}\ \emph {et~al.}(2016)\citenamefont {Welden},
  \citenamefont {Rusakov},\ and\ \citenamefont {Zgid}}]{welden2016statmech}%
  \BibitemOpen
  \bibfield  {author} {\bibinfo {author} {\bibfnamefont {A.~R.}\ \bibnamefont
  {Welden}}, \bibinfo {author} {\bibfnamefont {A.~A.}\ \bibnamefont
  {Rusakov}},\ and\ \bibinfo {author} {\bibfnamefont {D.}~\bibnamefont
  {Zgid}},\ }\bibfield  {title} {\bibinfo {title} {{Exploring connections
  between statistical mechanics and Green's functions for realistic systems:
  Temperature dependent electronic entropy and internal energy from a
  self-consistent second-order Green's function}},\ }\href
  {https://doi.org/10.1063/1.4967449} {\bibfield  {journal} {\bibinfo
  {journal} {The Journal of Chemical Physics}\ }\textbf {\bibinfo {volume}
  {145}},\ \bibinfo {pages} {204106} (\bibinfo {year} {2016})}\BibitemShut
  {NoStop}%
\bibitem [{\citenamefont {Iskakov}\ \emph {et~al.}(2019)\citenamefont
  {Iskakov}, \citenamefont {Rusakov}, \citenamefont {Zgid},\ and\ \citenamefont
  {Gull}}]{iskakov2018gf2}%
  \BibitemOpen
  \bibfield  {author} {\bibinfo {author} {\bibfnamefont {S.}~\bibnamefont
  {Iskakov}}, \bibinfo {author} {\bibfnamefont {A.~A.}\ \bibnamefont
  {Rusakov}}, \bibinfo {author} {\bibfnamefont {D.}~\bibnamefont {Zgid}},\ and\
  \bibinfo {author} {\bibfnamefont {E.}~\bibnamefont {Gull}},\ }\bibfield
  {title} {\bibinfo {title} {{Effect of propagator renormalization on the band
  gap of insulating solids}},\ }\href
  {https://doi.org/10.1103/PhysRevB.100.085112} {\bibfield  {journal} {\bibinfo
   {journal} {Physical Review B}\ }\textbf {\bibinfo {volume} {100}},\ \bibinfo
  {pages} {085112} (\bibinfo {year} {2019})}\BibitemShut {NoStop}%
\bibitem [{\citenamefont {Sun}\ and\ \citenamefont {Kotliar}(2002)}]{Sun2002}%
  \BibitemOpen
  \bibfield  {author} {\bibinfo {author} {\bibfnamefont {P.}~\bibnamefont
  {Sun}}\ and\ \bibinfo {author} {\bibfnamefont {G.}~\bibnamefont {Kotliar}},\
  }\bibfield  {title} {\bibinfo {title} {{Extended dynamical mean-field theory
  and $GW$ method}},\ }\href {https://doi.org/10.1103/PhysRevB.66.085120}
  {\bibfield  {journal} {\bibinfo  {journal} {Physical Review B}\ }\textbf
  {\bibinfo {volume} {66}},\ \bibinfo {pages} {085120} (\bibinfo {year}
  {2002})}\BibitemShut {NoStop}%
\bibitem [{\citenamefont {Biermann}\ \emph {et~al.}(2003)\citenamefont
  {Biermann}, \citenamefont {Aryasetiawan},\ and\ \citenamefont
  {Georges}}]{Biermann02}%
  \BibitemOpen
  \bibfield  {author} {\bibinfo {author} {\bibfnamefont {S.}~\bibnamefont
  {Biermann}}, \bibinfo {author} {\bibfnamefont {F.}~\bibnamefont
  {Aryasetiawan}},\ and\ \bibinfo {author} {\bibfnamefont {A.}~\bibnamefont
  {Georges}},\ }\bibfield  {title} {\bibinfo {title} {{First-Principles
  Approach to the Electronic Structure of Strongly Correlated Systems:
  Combining the $GW$ Approximation and Dynamical Mean-Field Theory}},\ }\href
  {https://doi.org/10.1103/PhysRevLett.90.086402} {\bibfield  {journal}
  {\bibinfo  {journal} {Physical Review Letters}\ }\textbf {\bibinfo {volume}
  {90}},\ \bibinfo {pages} {086402} (\bibinfo {year} {2003})}\BibitemShut
  {NoStop}%
\bibitem [{\citenamefont {Tomczak}\ \emph {et~al.}(2012)\citenamefont
  {Tomczak}, \citenamefont {Casula}, \citenamefont {Miyake}, \citenamefont
  {Aryasetiawan},\ and\ \citenamefont {Biermann}}]{Tomczak_2012}%
  \BibitemOpen
  \bibfield  {author} {\bibinfo {author} {\bibfnamefont {J.~M.}\ \bibnamefont
  {Tomczak}}, \bibinfo {author} {\bibfnamefont {M.}~\bibnamefont {Casula}},
  \bibinfo {author} {\bibfnamefont {T.}~\bibnamefont {Miyake}}, \bibinfo
  {author} {\bibfnamefont {F.}~\bibnamefont {Aryasetiawan}},\ and\ \bibinfo
  {author} {\bibfnamefont {S.}~\bibnamefont {Biermann}},\ }\bibfield  {title}
  {\bibinfo {title} {{Combined GW and dynamical mean-field theory: Dynamical
  screening effects in transition metal oxides}},\ }\href
  {https://doi.org/10.1209/0295-5075/100/67001} {\bibfield  {journal} {\bibinfo
   {journal} {EPL (Europhysics Letters)}\ }\textbf {\bibinfo {volume} {100}},\
  \bibinfo {pages} {67001} (\bibinfo {year} {2012})}\BibitemShut {NoStop}%
\bibitem [{\citenamefont {Tomczak}\ \emph {et~al.}(2017)\citenamefont
  {Tomczak}, \citenamefont {Liu}, \citenamefont {Toschi}, \citenamefont
  {Kresse},\ and\ \citenamefont {Held}}]{Tomczak2017}%
  \BibitemOpen
  \bibfield  {author} {\bibinfo {author} {\bibfnamefont {J.~M.}\ \bibnamefont
  {Tomczak}}, \bibinfo {author} {\bibfnamefont {P.}~\bibnamefont {Liu}},
  \bibinfo {author} {\bibfnamefont {A.}~\bibnamefont {Toschi}}, \bibinfo
  {author} {\bibfnamefont {G.}~\bibnamefont {Kresse}},\ and\ \bibinfo {author}
  {\bibfnamefont {K.}~\bibnamefont {Held}},\ }\bibfield  {title} {\bibinfo
  {title} {{Merging GW with DMFT and non-local correlations beyond}},\ }\href
  {https://doi.org/10.1140/epjst/e2017-70053-1} {\bibfield  {journal} {\bibinfo
   {journal} {European Physical Journal: Special Topics}\ }\textbf {\bibinfo
  {volume} {226}},\ \bibinfo {pages} {2565} (\bibinfo {year}
  {2017})}\BibitemShut {NoStop}%
\bibitem [{\citenamefont {Werner}\ and\ \citenamefont
  {Casula}(2016)}]{Werner2016}%
  \BibitemOpen
  \bibfield  {author} {\bibinfo {author} {\bibfnamefont {P.}~\bibnamefont
  {Werner}}\ and\ \bibinfo {author} {\bibfnamefont {M.}~\bibnamefont
  {Casula}},\ }\bibfield  {title} {\bibinfo {title} {{Dynamical screening in
  correlated electron systems—from lattice models to realistic materials}},\
  }\href {https://doi.org/10.1088/0953-8984/28/38/383001} {\bibfield  {journal}
  {\bibinfo  {journal} {Journal of Physics: Condensed Matter}\ }\textbf
  {\bibinfo {volume} {28}},\ \bibinfo {pages} {383001} (\bibinfo {year}
  {2016})}\BibitemShut {NoStop}%
\bibitem [{\citenamefont {Kananenka}\ \emph {et~al.}(2015)\citenamefont
  {Kananenka}, \citenamefont {Gull},\ and\ \citenamefont {Zgid}}]{Kananenka15}%
  \BibitemOpen
  \bibfield  {author} {\bibinfo {author} {\bibfnamefont {A.~A.}\ \bibnamefont
  {Kananenka}}, \bibinfo {author} {\bibfnamefont {E.}~\bibnamefont {Gull}},\
  and\ \bibinfo {author} {\bibfnamefont {D.}~\bibnamefont {Zgid}},\ }\bibfield
  {title} {\bibinfo {title} {{Systematically improvable multiscale solver for
  correlated electron systems}},\ }\href
  {https://doi.org/10.1103/PhysRevB.91.121111} {\bibfield  {journal} {\bibinfo
  {journal} {Physical Review B}\ }\textbf {\bibinfo {volume} {91}},\ \bibinfo
  {pages} {121111} (\bibinfo {year} {2015})}\BibitemShut {NoStop}%
\bibitem [{\citenamefont {Lan}\ \emph {et~al.}(2015)\citenamefont {Lan},
  \citenamefont {Kananenka},\ and\ \citenamefont
  {Zgid}}]{Lan2015communication}%
  \BibitemOpen
  \bibfield  {author} {\bibinfo {author} {\bibfnamefont {T.~N.}\ \bibnamefont
  {Lan}}, \bibinfo {author} {\bibfnamefont {A.~A.}\ \bibnamefont {Kananenka}},\
  and\ \bibinfo {author} {\bibfnamefont {D.}~\bibnamefont {Zgid}},\ }\bibfield
  {title} {\bibinfo {title} {{Communication: Towards ab initio self-energy
  embedding theory in quantum chemistry}},\ }\href
  {https://doi.org/10.1063/1.4938562} {\bibfield  {journal} {\bibinfo
  {journal} {The Journal of Chemical Physics}\ }\textbf {\bibinfo {volume}
  {143}},\ \bibinfo {pages} {241102} (\bibinfo {year} {2015})}\BibitemShut
  {NoStop}%
\bibitem [{\citenamefont {Zgid}\ and\ \citenamefont {Gull}(2017)}]{Zgid_2017}%
  \BibitemOpen
  \bibfield  {author} {\bibinfo {author} {\bibfnamefont {D.}~\bibnamefont
  {Zgid}}\ and\ \bibinfo {author} {\bibfnamefont {E.}~\bibnamefont {Gull}},\
  }\bibfield  {title} {\bibinfo {title} {{Finite temperature quantum embedding
  theories for correlated systems}},\ }\href
  {https://doi.org/10.1088/1367-2630/aa5d34} {\bibfield  {journal} {\bibinfo
  {journal} {New Journal of Physics}\ }\textbf {\bibinfo {volume} {19}},\
  \bibinfo {pages} {023047} (\bibinfo {year} {2017})}\BibitemShut {NoStop}%
\bibitem [{\citenamefont {Lan}\ and\ \citenamefont
  {Zgid}(2017)}]{Lan2017generalized}%
  \BibitemOpen
  \bibfield  {author} {\bibinfo {author} {\bibfnamefont {T.~N.}\ \bibnamefont
  {Lan}}\ and\ \bibinfo {author} {\bibfnamefont {D.}~\bibnamefont {Zgid}},\
  }\bibfield  {title} {\bibinfo {title} {{Generalized Self-Energy Embedding
  Theory}},\ }\href {https://doi.org/10.1021/acs.jpclett.7b00689} {\bibfield
  {journal} {\bibinfo  {journal} {The Journal of Physical Chemistry Letters}\
  }\textbf {\bibinfo {volume} {8}},\ \bibinfo {pages} {2200} (\bibinfo {year}
  {2017})}\BibitemShut {NoStop}%
\bibitem [{\citenamefont {Lan}\ \emph {et~al.}(2017)\citenamefont {Lan},
  \citenamefont {Shee}, \citenamefont {Li}, \citenamefont {Gull},\ and\
  \citenamefont {Zgid}}]{Lan:2017hk}%
  \BibitemOpen
  \bibfield  {author} {\bibinfo {author} {\bibfnamefont {T.~N.}\ \bibnamefont
  {Lan}}, \bibinfo {author} {\bibfnamefont {A.}~\bibnamefont {Shee}}, \bibinfo
  {author} {\bibfnamefont {J.}~\bibnamefont {Li}}, \bibinfo {author}
  {\bibfnamefont {E.}~\bibnamefont {Gull}},\ and\ \bibinfo {author}
  {\bibfnamefont {D.}~\bibnamefont {Zgid}},\ }\bibfield  {title} {\bibinfo
  {title} {{Testing self-energy embedding theory in combination with GW}},\
  }\href {https://doi.org/10.1103/PhysRevB.96.155106} {\bibfield  {journal}
  {\bibinfo  {journal} {Physical Review B}\ }\textbf {\bibinfo {volume} {96}},\
  \bibinfo {pages} {155106} (\bibinfo {year} {2017})}\BibitemShut {NoStop}%
\bibitem [{\citenamefont {Tran}\ \emph {et~al.}(2018)\citenamefont {Tran},
  \citenamefont {Iskakov},\ and\ \citenamefont
  {Zgid}}]{Tran2018spinunrestricted}%
  \BibitemOpen
  \bibfield  {author} {\bibinfo {author} {\bibfnamefont {L.~N.}\ \bibnamefont
  {Tran}}, \bibinfo {author} {\bibfnamefont {S.}~\bibnamefont {Iskakov}},\ and\
  \bibinfo {author} {\bibfnamefont {D.}~\bibnamefont {Zgid}},\ }\bibfield
  {title} {\bibinfo {title} {{Spin-Unrestricted Self-Energy Embedding
  Theory}},\ }\href {https://doi.org/10.1021/acs.jpclett.8b01754} {\bibfield
  {journal} {\bibinfo  {journal} {The Journal of Physical Chemistry Letters}\
  }\textbf {\bibinfo {volume} {9}},\ \bibinfo {pages} {4444} (\bibinfo {year}
  {2018})}\BibitemShut {NoStop}%
\bibitem [{\citenamefont {Rusakov}\ \emph {et~al.}(2019)\citenamefont
  {Rusakov}, \citenamefont {Iskakov}, \citenamefont {Tran},\ and\ \citenamefont
  {Zgid}}]{Rusakov2019seet}%
  \BibitemOpen
  \bibfield  {author} {\bibinfo {author} {\bibfnamefont {A.~A.}\ \bibnamefont
  {Rusakov}}, \bibinfo {author} {\bibfnamefont {S.}~\bibnamefont {Iskakov}},
  \bibinfo {author} {\bibfnamefont {L.~N.}\ \bibnamefont {Tran}},\ and\
  \bibinfo {author} {\bibfnamefont {D.}~\bibnamefont {Zgid}},\ }\bibfield
  {title} {\bibinfo {title} {{Self-Energy Embedding Theory (SEET) for Periodic
  Systems}},\ }\href {https://doi.org/10.1021/acs.jctc.8b00927} {\bibfield
  {journal} {\bibinfo  {journal} {Journal of Chemical Theory and Computation}\
  }\textbf {\bibinfo {volume} {15}},\ \bibinfo {pages} {229} (\bibinfo {year}
  {2019})}\BibitemShut {NoStop}%
\bibitem [{\citenamefont {Ku}(2000)}]{ku2000thesis}%
  \BibitemOpen
  \bibfield  {author} {\bibinfo {author} {\bibfnamefont {W.}~\bibnamefont
  {Ku}},\ }\emph {\bibinfo {title} {{Electronic Excitations in Metals and
  Semiconductors: Ab Initio Studies of Realistic Many-Particle Systems}}},\
  \href {http://trace.tennessee.edu/utk_graddiss/2030} {Ph.D. thesis},\
  \bibinfo  {school} {University of Tennessee} (\bibinfo {year}
  {2000})\BibitemShut {NoStop}%
\bibitem [{\citenamefont {Ku}\ and\ \citenamefont {Eguiluz}(2002)}]{Ku2002}%
  \BibitemOpen
  \bibfield  {author} {\bibinfo {author} {\bibfnamefont {W.}~\bibnamefont
  {Ku}}\ and\ \bibinfo {author} {\bibfnamefont {A.~G.}\ \bibnamefont
  {Eguiluz}},\ }\bibfield  {title} {\bibinfo {title} {{Band-Gap Problem in
  Semiconductors Revisited: Effects of Core States and Many-Body
  Self-Consistency}},\ }\href {https://doi.org/10.1103/PhysRevLett.89.126401}
  {\bibfield  {journal} {\bibinfo  {journal} {Physical Review Letters}\
  }\textbf {\bibinfo {volume} {89}},\ \bibinfo {pages} {126401} (\bibinfo
  {year} {2002})}\BibitemShut {NoStop}%
\bibitem [{\citenamefont {Boehnke}\ \emph {et~al.}(2011)\citenamefont
  {Boehnke}, \citenamefont {Hafermann}, \citenamefont {Ferrero}, \citenamefont
  {Lechermann},\ and\ \citenamefont {Parcollet}}]{boehnke2011legendre}%
  \BibitemOpen
  \bibfield  {author} {\bibinfo {author} {\bibfnamefont {L.}~\bibnamefont
  {Boehnke}}, \bibinfo {author} {\bibfnamefont {H.}~\bibnamefont {Hafermann}},
  \bibinfo {author} {\bibfnamefont {M.}~\bibnamefont {Ferrero}}, \bibinfo
  {author} {\bibfnamefont {F.}~\bibnamefont {Lechermann}},\ and\ \bibinfo
  {author} {\bibfnamefont {O.}~\bibnamefont {Parcollet}},\ }\bibfield  {title}
  {\bibinfo {title} {{Orthogonal polynomial representation of imaginary-time
  Green's functions}},\ }\href {https://doi.org/10.1103/PhysRevB.84.075145}
  {\bibfield  {journal} {\bibinfo  {journal} {Physical Review B}\ }\textbf
  {\bibinfo {volume} {84}},\ \bibinfo {pages} {075145} (\bibinfo {year}
  {2011})}\BibitemShut {NoStop}%
\bibitem [{\citenamefont {Gull}\ \emph {et~al.}(2018)\citenamefont {Gull},
  \citenamefont {Iskakov}, \citenamefont {Krivenko}, \citenamefont {Rusakov},\
  and\ \citenamefont {Zgid}}]{gull2018chebyshev}%
  \BibitemOpen
  \bibfield  {author} {\bibinfo {author} {\bibfnamefont {E.}~\bibnamefont
  {Gull}}, \bibinfo {author} {\bibfnamefont {S.}~\bibnamefont {Iskakov}},
  \bibinfo {author} {\bibfnamefont {I.}~\bibnamefont {Krivenko}}, \bibinfo
  {author} {\bibfnamefont {A.~A.}\ \bibnamefont {Rusakov}},\ and\ \bibinfo
  {author} {\bibfnamefont {D.}~\bibnamefont {Zgid}},\ }\bibfield  {title}
  {\bibinfo {title} {{Chebyshev polynomial representation of imaginary-time
  response functions}},\ }\href {https://doi.org/10.1103/PhysRevB.98.075127}
  {\bibfield  {journal} {\bibinfo  {journal} {Physical Review B}\ }\textbf
  {\bibinfo {volume} {98}},\ \bibinfo {pages} {75127} (\bibinfo {year}
  {2018})}\BibitemShut {NoStop}%
\bibitem [{\citenamefont {Kutepov}\ \emph {et~al.}(2012)\citenamefont
  {Kutepov}, \citenamefont {Haule}, \citenamefont {Savrasov},\ and\
  \citenamefont {Kotliar}}]{kutepov2012electronic}%
  \BibitemOpen
  \bibfield  {author} {\bibinfo {author} {\bibfnamefont {A.}~\bibnamefont
  {Kutepov}}, \bibinfo {author} {\bibfnamefont {K.}~\bibnamefont {Haule}},
  \bibinfo {author} {\bibfnamefont {S.~Y.}\ \bibnamefont {Savrasov}},\ and\
  \bibinfo {author} {\bibfnamefont {G.}~\bibnamefont {Kotliar}},\ }\bibfield
  {title} {\bibinfo {title} {{Electronic structure of Pu and Am metals by
  self-consistent relativistic $GW$ method}},\ }\href
  {https://doi.org/10.1103/PhysRevB.85.155129} {\bibfield  {journal} {\bibinfo
  {journal} {Physical Review B}\ }\textbf {\bibinfo {volume} {85}},\ \bibinfo
  {pages} {155129} (\bibinfo {year} {2012})}\BibitemShut {NoStop}%
\bibitem [{\citenamefont {Shinaoka}\ \emph {et~al.}(2017)\citenamefont
  {Shinaoka}, \citenamefont {Otsuki}, \citenamefont {Ohzeki},\ and\
  \citenamefont {Yoshimi}}]{shinaoka2017compressing}%
  \BibitemOpen
  \bibfield  {author} {\bibinfo {author} {\bibfnamefont {H.}~\bibnamefont
  {Shinaoka}}, \bibinfo {author} {\bibfnamefont {J.}~\bibnamefont {Otsuki}},
  \bibinfo {author} {\bibfnamefont {M.}~\bibnamefont {Ohzeki}},\ and\ \bibinfo
  {author} {\bibfnamefont {K.}~\bibnamefont {Yoshimi}},\ }\bibfield  {title}
  {\bibinfo {title} {{Compressing Green's function using intermediate
  representation between imaginary-time and real-frequency domains}},\ }\href
  {https://doi.org/10.1103/PhysRevB.96.035147} {\bibfield  {journal} {\bibinfo
  {journal} {Physical Review B}\ }\textbf {\bibinfo {volume} {96}},\ \bibinfo
  {pages} {35147} (\bibinfo {year} {2017})}\BibitemShut {NoStop}%
\bibitem [{\citenamefont {Chikano}\ \emph {et~al.}(2018)\citenamefont
  {Chikano}, \citenamefont {Otsuki},\ and\ \citenamefont
  {Shinaoka}}]{chikano2018performance}%
  \BibitemOpen
  \bibfield  {author} {\bibinfo {author} {\bibfnamefont {N.}~\bibnamefont
  {Chikano}}, \bibinfo {author} {\bibfnamefont {J.}~\bibnamefont {Otsuki}},\
  and\ \bibinfo {author} {\bibfnamefont {H.}~\bibnamefont {Shinaoka}},\
  }\bibfield  {title} {\bibinfo {title} {{Performance analysis of a physically
  constructed orthogonal representation of imaginary-time Green's function}},\
  }\href {https://doi.org/10.1103/PhysRevB.98.035104} {\bibfield  {journal}
  {\bibinfo  {journal} {Physical Review B}\ }\textbf {\bibinfo {volume} {98}},\
  \bibinfo {pages} {35104} (\bibinfo {year} {2018})}\BibitemShut {NoStop}%
\bibitem [{\citenamefont {Chikano}\ \emph {et~al.}(2019)\citenamefont
  {Chikano}, \citenamefont {Yoshimi}, \citenamefont {Otsuki},\ and\
  \citenamefont {Shinaoka}}]{chikano2019irbasis}%
  \BibitemOpen
  \bibfield  {author} {\bibinfo {author} {\bibfnamefont {N.}~\bibnamefont
  {Chikano}}, \bibinfo {author} {\bibfnamefont {K.}~\bibnamefont {Yoshimi}},
  \bibinfo {author} {\bibfnamefont {J.}~\bibnamefont {Otsuki}},\ and\ \bibinfo
  {author} {\bibfnamefont {H.}~\bibnamefont {Shinaoka}},\ }\bibfield  {title}
  {\bibinfo {title} {{irbasis: Open-source database and software for
  intermediate-representation basis functions of imaginary-time Green's
  function}},\ }\href {https://doi.org/10.1016/j.cpc.2019.02.006} {\bibfield
  {journal} {\bibinfo  {journal} {Computer Physics Communications}\ }\textbf
  {\bibinfo {volume} {240}},\ \bibinfo {pages} {181} (\bibinfo {year}
  {2019})}\BibitemShut {NoStop}%
\bibitem [{\citenamefont {Steinbeck}\ \emph {et~al.}(2000)\citenamefont
  {Steinbeck}, \citenamefont {Rubio}, \citenamefont {Reining}, \citenamefont
  {Torrent}, \citenamefont {White},\ and\ \citenamefont
  {Godby}}]{steinbeck2000enhancements}%
  \BibitemOpen
  \bibfield  {author} {\bibinfo {author} {\bibfnamefont {L.}~\bibnamefont
  {Steinbeck}}, \bibinfo {author} {\bibfnamefont {A.}~\bibnamefont {Rubio}},
  \bibinfo {author} {\bibfnamefont {L.}~\bibnamefont {Reining}}, \bibinfo
  {author} {\bibfnamefont {M.}~\bibnamefont {Torrent}}, \bibinfo {author}
  {\bibfnamefont {I.}~\bibnamefont {White}},\ and\ \bibinfo {author}
  {\bibfnamefont {R.}~\bibnamefont {Godby}},\ }\bibfield  {title} {\bibinfo
  {title} {{Enhancements to the GW space-time method}},\ }\href
  {https://doi.org/10.1016/S0010-4655(99)00466-X} {\bibfield  {journal}
  {\bibinfo  {journal} {Computer Physics Communications}\ }\textbf {\bibinfo
  {volume} {125}},\ \bibinfo {pages} {105} (\bibinfo {year}
  {2000})}\BibitemShut {NoStop}%
\bibitem [{\citenamefont {Kaltak}\ \emph {et~al.}(2014)\citenamefont {Kaltak},
  \citenamefont {Klime{\v{s}}},\ and\ \citenamefont {Kresse}}]{kaltak2014low}%
  \BibitemOpen
  \bibfield  {author} {\bibinfo {author} {\bibfnamefont {M.}~\bibnamefont
  {Kaltak}}, \bibinfo {author} {\bibfnamefont {J.}~\bibnamefont
  {Klime{\v{s}}}},\ and\ \bibinfo {author} {\bibfnamefont {G.}~\bibnamefont
  {Kresse}},\ }\bibfield  {title} {\bibinfo {title} {{Low scaling algorithms
  for the random phase approximation: Imaginary time and laplace
  transformations}},\ }\href {https://doi.org/10.1021/ct5001268} {\bibfield
  {journal} {\bibinfo  {journal} {Journal of Chemical Theory and Computation}\
  }\textbf {\bibinfo {volume} {10}},\ \bibinfo {pages} {2498} (\bibinfo {year}
  {2014})}\BibitemShut {NoStop}%
\bibitem [{\citenamefont {Rusakov}\ \emph {et~al.}(2014)\citenamefont
  {Rusakov}, \citenamefont {Phillips},\ and\ \citenamefont
  {Zgid}}]{Rusakov2014}%
  \BibitemOpen
  \bibfield  {author} {\bibinfo {author} {\bibfnamefont {A.~A.}\ \bibnamefont
  {Rusakov}}, \bibinfo {author} {\bibfnamefont {J.~J.}\ \bibnamefont
  {Phillips}},\ and\ \bibinfo {author} {\bibfnamefont {D.}~\bibnamefont
  {Zgid}},\ }\bibfield  {title} {\bibinfo {title} {{Local Hamiltonians for
  quantitative Green's function embedding methods}},\ }\href
  {https://doi.org/10.1063/1.4901432} {\bibfield  {journal} {\bibinfo
  {journal} {Journal of Chemical Physics}\ }\textbf {\bibinfo {volume} {141}},\
  \bibinfo {pages} {194105} (\bibinfo {year} {2014})}\BibitemShut {NoStop}%
\bibitem [{\citenamefont {Comanac}(2007)}]{ArminPhD}%
  \BibitemOpen
  \bibfield  {author} {\bibinfo {author} {\bibfnamefont {A.-B.}\ \bibnamefont
  {Comanac}},\ }\emph {\bibinfo {title} {{Dynamical Mean Field Theroy of
  Correlated Electron Systems: New Algorithms and Applications to Local
  Observables}}},\ \href@noop {} {Ph.D. thesis},\ \bibinfo  {school} {Columbia
  University} (\bibinfo {year} {2007})\BibitemShut {NoStop}%
\bibitem [{\citenamefont {Bl{\"{u}}mer}(2002)}]{BluemerPhD}%
  \BibitemOpen
  \bibfield  {author} {\bibinfo {author} {\bibfnamefont {N.}~\bibnamefont
  {Bl{\"{u}}mer}},\ }\emph {\bibinfo {title} {{Mott-Hubbard Metal-Insulator
  Transition and Optical Conductivity in High Dimensions}}},\ \href
  {http://komet337.physik.uni-mainz.de/Bluemer/thesis%5Cnhttp://www.shaker.de/de/content/catalogue/index.asp?lang=de&ID=8&ISBN=978-3-8322-2320-5}
  {Ph.D. thesis},\ \bibinfo  {school} {Universit{\"{a}}t Augsburg} (\bibinfo
  {year} {2002})\BibitemShut {NoStop}%
\bibitem [{\citenamefont {Szabo}\ and\ \citenamefont
  {Ostlund}(1989)}]{ostlund1996modern}%
  \BibitemOpen
  \bibfield  {author} {\bibinfo {author} {\bibfnamefont {A.}~\bibnamefont
  {Szabo}}\ and\ \bibinfo {author} {\bibfnamefont {N.~S.}\ \bibnamefont
  {Ostlund}},\ }\href@noop {} {\emph {\bibinfo {title} {McGraw-Hili, New
  York}}}\ (\bibinfo  {publisher} {Dover Publications Inc New edition edn},\
  \bibinfo {year} {1989})\BibitemShut {NoStop}%
\bibitem [{\citenamefont {Bohm}\ and\ \citenamefont
  {Pines}(1953)}]{PhysRev.92.609}%
  \BibitemOpen
  \bibfield  {author} {\bibinfo {author} {\bibfnamefont {D.}~\bibnamefont
  {Bohm}}\ and\ \bibinfo {author} {\bibfnamefont {D.}~\bibnamefont {Pines}},\
  }\bibfield  {title} {\bibinfo {title} {{A Collective Description of Electron
  Interactions: III. Coulomb Interactions in a Degenerate Electron Gas}},\
  }\href {https://doi.org/10.1103/PhysRev.92.609} {\bibfield  {journal}
  {\bibinfo  {journal} {Physical Review}\ }\textbf {\bibinfo {volume} {92}},\
  \bibinfo {pages} {609} (\bibinfo {year} {1953})}\BibitemShut {NoStop}%
\bibitem [{\citenamefont {Motta}\ \emph {et~al.}(2017)\citenamefont {Motta},
  \citenamefont {Ceperley}, \citenamefont {Chan}, \citenamefont {Gomez},
  \citenamefont {Gull}, \citenamefont {Guo}, \citenamefont
  {Jim{\'{e}}nez-Hoyos}, \citenamefont {Lan}, \citenamefont {Li}, \citenamefont
  {Ma}, \citenamefont {Millis}, \citenamefont {Prokof'ev}, \citenamefont {Ray},
  \citenamefont {Scuseria}, \citenamefont {Sorella}, \citenamefont
  {Stoudenmire}, \citenamefont {Sun}, \citenamefont {Tupitsyn}, \citenamefont
  {White}, \citenamefont {Zgid},\ and\ \citenamefont
  {Zhang}}]{motta2017hydrogen}%
  \BibitemOpen
  \bibfield  {author} {\bibinfo {author} {\bibfnamefont {M.}~\bibnamefont
  {Motta}}, \bibinfo {author} {\bibfnamefont {D.~M.}\ \bibnamefont {Ceperley}},
  \bibinfo {author} {\bibfnamefont {G.~K.-L.}\ \bibnamefont {Chan}}, \bibinfo
  {author} {\bibfnamefont {J.~A.}\ \bibnamefont {Gomez}}, \bibinfo {author}
  {\bibfnamefont {E.}~\bibnamefont {Gull}}, \bibinfo {author} {\bibfnamefont
  {S.}~\bibnamefont {Guo}}, \bibinfo {author} {\bibfnamefont {C.~A.}\
  \bibnamefont {Jim{\'{e}}nez-Hoyos}}, \bibinfo {author} {\bibfnamefont
  {T.~N.}\ \bibnamefont {Lan}}, \bibinfo {author} {\bibfnamefont
  {J.}~\bibnamefont {Li}}, \bibinfo {author} {\bibfnamefont {F.}~\bibnamefont
  {Ma}}, \bibinfo {author} {\bibfnamefont {A.~J.}\ \bibnamefont {Millis}},
  \bibinfo {author} {\bibfnamefont {N.~V.}\ \bibnamefont {Prokof'ev}}, \bibinfo
  {author} {\bibfnamefont {U.}~\bibnamefont {Ray}}, \bibinfo {author}
  {\bibfnamefont {G.~E.}\ \bibnamefont {Scuseria}}, \bibinfo {author}
  {\bibfnamefont {S.}~\bibnamefont {Sorella}}, \bibinfo {author} {\bibfnamefont
  {E.~M.}\ \bibnamefont {Stoudenmire}}, \bibinfo {author} {\bibfnamefont
  {Q.}~\bibnamefont {Sun}}, \bibinfo {author} {\bibfnamefont {I.~S.}\
  \bibnamefont {Tupitsyn}}, \bibinfo {author} {\bibfnamefont {S.~R.}\
  \bibnamefont {White}}, \bibinfo {author} {\bibfnamefont {D.}~\bibnamefont
  {Zgid}},\ and\ \bibinfo {author} {\bibfnamefont {S.}~\bibnamefont {Zhang}},\
  }\bibfield  {title} {\bibinfo {title} {{Towards the Solution of the
  Many-Electron Problem in Real Materials: Equation of State of the Hydrogen
  Chain with State-of-the-Art Many-Body Methods}},\ }\href
  {https://doi.org/10.1103/PhysRevX.7.031059} {\bibfield  {journal} {\bibinfo
  {journal} {Physical Review X}\ }\textbf {\bibinfo {volume} {7}},\ \bibinfo
  {pages} {031059} (\bibinfo {year} {2017})}\BibitemShut {NoStop}%
\bibitem [{\citenamefont {Sun}\ \emph {et~al.}(2018)\citenamefont {Sun},
  \citenamefont {Berkelbach}, \citenamefont {Blunt}, \citenamefont {Booth},
  \citenamefont {Guo}, \citenamefont {Li}, \citenamefont {Liu}, \citenamefont
  {McClain}, \citenamefont {Sayfutyarova}, \citenamefont {Sharma},
  \citenamefont {Wouters},\ and\ \citenamefont {Chan}}]{PYSCF}%
  \BibitemOpen
  \bibfield  {author} {\bibinfo {author} {\bibfnamefont {Q.}~\bibnamefont
  {Sun}}, \bibinfo {author} {\bibfnamefont {T.~C.}\ \bibnamefont {Berkelbach}},
  \bibinfo {author} {\bibfnamefont {N.~S.}\ \bibnamefont {Blunt}}, \bibinfo
  {author} {\bibfnamefont {G.~H.}\ \bibnamefont {Booth}}, \bibinfo {author}
  {\bibfnamefont {S.}~\bibnamefont {Guo}}, \bibinfo {author} {\bibfnamefont
  {Z.}~\bibnamefont {Li}}, \bibinfo {author} {\bibfnamefont {J.}~\bibnamefont
  {Liu}}, \bibinfo {author} {\bibfnamefont {J.~D.}\ \bibnamefont {McClain}},
  \bibinfo {author} {\bibfnamefont {E.~R.}\ \bibnamefont {Sayfutyarova}},
  \bibinfo {author} {\bibfnamefont {S.}~\bibnamefont {Sharma}}, \bibinfo
  {author} {\bibfnamefont {S.}~\bibnamefont {Wouters}},\ and\ \bibinfo {author}
  {\bibfnamefont {G.~K.~L.}\ \bibnamefont {Chan}},\ }\bibfield  {title}
  {\bibinfo {title} {{PySCF: the Python-based simulations of chemistry
  framework}},\ }\href {https://doi.org/10.1002/wcms.1340} {\bibfield
  {journal} {\bibinfo  {journal} {Wiley Interdisciplinary Reviews:
  Computational Molecular Science}\ }\textbf {\bibinfo {volume} {8}},\ \bibinfo
  {pages} {e1340} (\bibinfo {year} {2018})}\BibitemShut {NoStop}%
\bibitem [{\citenamefont {Dunning}(1989)}]{Dunning1989}%
  \BibitemOpen
  \bibfield  {author} {\bibinfo {author} {\bibfnamefont {T.~H.}\ \bibnamefont
  {Dunning}},\ }\bibfield  {title} {\bibinfo {title} {{Gaussian basis sets for
  use in correlated molecular calculations. I. The atoms boron through neon and
  hydrogen}},\ }\href {https://doi.org/10.1063/1.456153} {\bibfield  {journal}
  {\bibinfo  {journal} {The Journal of Chemical Physics}\ }\textbf {\bibinfo
  {volume} {90}},\ \bibinfo {pages} {1007} (\bibinfo {year}
  {1989})}\BibitemShut {NoStop}%
\bibitem [{\citenamefont {Goedecker}\ \emph {et~al.}(1996)\citenamefont
  {Goedecker}, \citenamefont {Teter},\ and\ \citenamefont
  {Hutter}}]{Goedecker1996}%
  \BibitemOpen
  \bibfield  {author} {\bibinfo {author} {\bibfnamefont {S.}~\bibnamefont
  {Goedecker}}, \bibinfo {author} {\bibfnamefont {M.}~\bibnamefont {Teter}},\
  and\ \bibinfo {author} {\bibfnamefont {J.}~\bibnamefont {Hutter}},\
  }\bibfield  {title} {\bibinfo {title} {{Separable dual-space Gaussian
  pseudopotentials}},\ }\href {https://doi.org/10.1103/PhysRevB.54.1703}
  {\bibfield  {journal} {\bibinfo  {journal} {Physical Review B}\ }\textbf
  {\bibinfo {volume} {54}},\ \bibinfo {pages} {1703} (\bibinfo {year}
  {1996})}\BibitemShut {NoStop}%
\bibitem [{\citenamefont {Sano}\ \emph {et~al.}(2016)\citenamefont {Sano},
  \citenamefont {Koretsune}, \citenamefont {Tadano}, \citenamefont {Akashi},\
  and\ \citenamefont {Arita}}]{sano2016eliashberg}%
  \BibitemOpen
  \bibfield  {author} {\bibinfo {author} {\bibfnamefont {W.}~\bibnamefont
  {Sano}}, \bibinfo {author} {\bibfnamefont {T.}~\bibnamefont {Koretsune}},
  \bibinfo {author} {\bibfnamefont {T.}~\bibnamefont {Tadano}}, \bibinfo
  {author} {\bibfnamefont {R.}~\bibnamefont {Akashi}},\ and\ \bibinfo {author}
  {\bibfnamefont {R.}~\bibnamefont {Arita}},\ }\bibfield  {title} {\bibinfo
  {title} {{Effect of Van Hove singularities on high-$T_\mathrm{c}$
  superconductivity in $\mathrm{H}_3\mathrm{S}$}},\ }\href
  {https://doi.org/10.1103/PhysRevB.93.094525} {\bibfield  {journal} {\bibinfo
  {journal} {Physical Review B}\ }\textbf {\bibinfo {volume} {93}},\ \bibinfo
  {pages} {094525} (\bibinfo {year} {2016})}\BibitemShut {NoStop}%
\bibitem [{\citenamefont {Jarrell}(1992)}]{Jarrell92}%
  \BibitemOpen
  \bibfield  {author} {\bibinfo {author} {\bibfnamefont {M.}~\bibnamefont
  {Jarrell}},\ }\bibfield  {title} {\bibinfo {title} {{Hubbard model in
  infinite dimensions: A quantum Monte Carlo study}},\ }\href
  {https://doi.org/10.1103/PhysRevLett.69.168} {\bibfield  {journal} {\bibinfo
  {journal} {Physical Review Letters}\ }\textbf {\bibinfo {volume} {69}},\
  \bibinfo {pages} {168} (\bibinfo {year} {1992})}\BibitemShut {NoStop}%
\bibitem [{\citenamefont {Held}\ \emph {et~al.}(2009)\citenamefont {Held},
  \citenamefont {Katanin},\ and\ \citenamefont {Toschi}}]{Karsten2008}%
  \BibitemOpen
  \bibfield  {author} {\bibinfo {author} {\bibfnamefont {K.}~\bibnamefont
  {Held}}, \bibinfo {author} {\bibfnamefont {A.~A.}\ \bibnamefont {Katanin}},\
  and\ \bibinfo {author} {\bibfnamefont {A.}~\bibnamefont {Toschi}},\
  }\bibfield  {title} {\bibinfo {title} {{Dynamical Vertex Approximation}},\
  }\href {https://doi.org/10.1143/ptps.176.117} {\bibfield  {journal} {\bibinfo
   {journal} {Progress of Theoretical Physics Supplement}\ }\textbf {\bibinfo
  {volume} {176}},\ \bibinfo {pages} {117} (\bibinfo {year}
  {2009})}\BibitemShut {NoStop}%
\bibitem [{\citenamefont {Shinaoka}\ \emph {et~al.}(2018)\citenamefont
  {Shinaoka}, \citenamefont {Otsuki}, \citenamefont {Haule}, \citenamefont
  {Wallerberger}, \citenamefont {Gull}, \citenamefont {Yoshimi},\ and\
  \citenamefont {Ohzeki}}]{Shinaoka:2018cg}%
  \BibitemOpen
  \bibfield  {author} {\bibinfo {author} {\bibfnamefont {H.}~\bibnamefont
  {Shinaoka}}, \bibinfo {author} {\bibfnamefont {J.}~\bibnamefont {Otsuki}},
  \bibinfo {author} {\bibfnamefont {K.}~\bibnamefont {Haule}}, \bibinfo
  {author} {\bibfnamefont {M.}~\bibnamefont {Wallerberger}}, \bibinfo {author}
  {\bibfnamefont {E.}~\bibnamefont {Gull}}, \bibinfo {author} {\bibfnamefont
  {K.}~\bibnamefont {Yoshimi}},\ and\ \bibinfo {author} {\bibfnamefont
  {M.}~\bibnamefont {Ohzeki}},\ }\bibfield  {title} {\bibinfo {title}
  {{Overcomplete compact representation of two-particle Green's functions}},\
  }\href {https://doi.org/10.1103/PhysRevB.97.205111} {\bibfield  {journal}
  {\bibinfo  {journal} {Physical Review B}\ }\textbf {\bibinfo {volume} {97}},\
  \bibinfo {pages} {205111} (\bibinfo {year} {2018})}\BibitemShut {NoStop}%
\bibitem [{\citenamefont {Otsuki}\ \emph {et~al.}(2017)\citenamefont {Otsuki},
  \citenamefont {Ohzeki}, \citenamefont {Shinaoka},\ and\ \citenamefont
  {Yoshimi}}]{Otsuki:2017er}%
  \BibitemOpen
  \bibfield  {author} {\bibinfo {author} {\bibfnamefont {J.}~\bibnamefont
  {Otsuki}}, \bibinfo {author} {\bibfnamefont {M.}~\bibnamefont {Ohzeki}},
  \bibinfo {author} {\bibfnamefont {H.}~\bibnamefont {Shinaoka}},\ and\
  \bibinfo {author} {\bibfnamefont {K.}~\bibnamefont {Yoshimi}},\ }\bibfield
  {title} {\bibinfo {title} {{Sparse modeling approach to analytical
  continuation of imaginary-time quantum Monte Carlo data}},\ }\href
  {https://doi.org/10.1103/PhysRevE.95.061302} {\bibfield  {journal} {\bibinfo
  {journal} {Physical Review E}\ }\textbf {\bibinfo {volume} {95}},\ \bibinfo
  {pages} {061302(R)} (\bibinfo {year} {2017})}\BibitemShut {NoStop}%
\bibitem [{\citenamefont {Nagai}\ and\ \citenamefont
  {Shinaoka}(2019)}]{Nagai:2019dea}%
  \BibitemOpen
  \bibfield  {author} {\bibinfo {author} {\bibfnamefont {Y.}~\bibnamefont
  {Nagai}}\ and\ \bibinfo {author} {\bibfnamefont {H.}~\bibnamefont
  {Shinaoka}},\ }\bibfield  {title} {\bibinfo {title} {{Smooth Self-energy in
  the Exact-diagonalization-based Dynamical Mean-field Theory:
  Intermediate-representation Filtering Approach}},\ }\href
  {https://doi.org/10.7566/JPSJ.88.064004} {\bibfield  {journal} {\bibinfo
  {journal} {Journal of the Physical Society of Japan}\ }\textbf {\bibinfo
  {volume} {88}},\ \bibinfo {pages} {064004} (\bibinfo {year}
  {2019})}\BibitemShut {NoStop}%
\end{thebibliography}%

\end{document}